\documentclass[12pt]{article}

\usepackage{graphicx}
\usepackage{amssymb,amsmath,cite,color}
\newcommand{\degr}{\mbox{$^{\circ}$}}
\begin{document}
\parindent4mm

\title{ Intermediate Mass Fragment Emission in ${^{32}}$S
+${^{51}}$V, ${^{109}}$Ag,  and
${^{238}}$U Collisions at E~=~31.6~MeV~A}

\author{{\em{H. Machner }}
\\Fachbereich Physik, Univ. Duisburg-Essen
\\ Lotharstr. 1, 47048 Duisburg, Germany
\\  \\{\em{M. Nolte, M. Palarczyk}}
\\Institut f\"{u}r Kernphysik, Forschungszentrum J\"{u}lich,
\\52425 J\"{u}lich, Germany
\\ \\ {\em{T. Kutsarova}}
\\Institute of Nuclear Physics and Nuclear Energy, Sofia, Bulgaria}

\date{}

\maketitle

\begin{abstract}
Intermediate mass fragment emission  for  reactions
 of
$^{32}\text{S} +\,^{51}\text{V},\,^{109}$Ag,  and $^{238}\text{U}$ has  been  studied.
Double differential
cross sections were analysed in terms of the generalised moving source
model yielding charge distributions.
Isotope ratios show strong fragment mass dependencies.The data were successfully reproduced by the coalescence model as well as by statistical multifragmentation model calculations. Quantum molecular dynamics model calculations were not so successful.
\end{abstract}

\section{Introduction}\label{sec:Introduction}

The energy range from 10 to 100 MeV A is believed to be  a transitional one because of the decreasing importance  of  the  Pauli principle  with increasing excitation energy. The study of energetic light particles has been proven to be a good  tool to  investigate non-equilibrium properties in heavy ion reactions ~\cite{Holub86}, \cite{Mac83}, \cite{Mac85}. However,  heavy  ion reactions may proceed through rather complex processes which may be studied through intermediate mass fragment (IMF) emission $(Z>2)$. Complex fragment production is often assumed to occur via coalescence of nucleons which are close in phase space. Applications of the model at relativistic energies are in Refs. \cite{Gutbrod76}, \cite{Gosset77}, and \cite{Lemaire79}. Examples for energies of a few tens of MeV are in Refs. \cite{Awes80}, \cite{Datta87} and \cite{Hagel00}. The original formulation of the model goes back to Ref. \cite{Schwarz63}. Derivations in the context of heavy ion reactions are in \cite{Mekjian77} and \cite{Kapusta80}. Sato and Yazaki \cite{Sato81} derived a model where the size of the fragments is explicitly included via their wave functions. The structure of the formulae looks like the coalescence model, although no thermal equilibrium is required. IMF's can also be considered to be emitted through a phase transition from liquid to vapour ~\cite{Hirsch84,Lop84,Mac85,Gro82} which may be reached in the expansion phase of the hot fused system~\cite{Ber83}. The yield within this model depends in the grand-canonical limit on a power law times an exponential. The power law is $A^\tau$ with $A$ the mass number and $\tau$ the critical exponent. The exponent consists of the free energy plus the proton and neutron chemical potential divided by the temperature. For the free energy a liquid drop expansion is usually assumed \cite{Hirsch84}, \cite{Mac85}. This is also true for the statistical multifragmentation model (SMM) \cite{Bondorf}. Results of this model were recently compared to fragment spectra from proton induced reactions in the GeV area \cite{Fidelus14}. The liquid drop expansion contains the symmetry energy which is also part of the equation of state which is important for many astrophysical cases esp. properties of neutron stars. Because of this important connection a lot of recent studies deal with the symmetry energy in fragmentation reactions \cite{Buyukcizmeki12}, \cite{Soisson12}, \cite{ma13}, \cite{Mallik13}, \cite{Marini:2012}. Other work concentrates more on the target $N/Z$ dependence \cite{Laforest99}, \cite{Liu:2012},\cite{Lombardo12}.

A review of earlier models and  data is given in ~\cite{Huf85, Lyn87}  and more recent work is reviewed in \cite{Borderie08, Pochodzalla97, Planeta06}. An approach making use of many-body Green's functions theory is given in \cite{Muether08}. On  the  other  hand a successful analysis of IMF energy spectra in terms of a generalised moving source model indicates that one general mechanism might be responsible for light particle emission and IMF ~\cite{Mach90, Buh92, Mac06} emission as well. This approach works with proton induced reactions as well as in heavy ion reactions. Charge correlations for IMF emission for the present systems as well as for oxygen induced reactions at the same beam energy per nucleon are recently published \cite{Rama08, Mac08}. These papers contain details of the experiments and also studies of the underlying reaction mechanism. In this work we limit ourselves to inclusive data from sulphur-induced reactions at 32 MeV A measured for targets ranging from vanadium to uranium. After a short description of the experiments we will apply the moving source model to extract total cross sections. Then the question of isospin influence is discussed \cite{Marini:2012}.

\section{ Experiments}\label{sec:Experiments}

The experiments were carried out at  the J\"{u}lich  isochronous cyclotron which accelerated $^{32}\text{S}^{13+}$ ions to 1.011 GeV. Typical beam currents were 0.6 to 3 particle nA depending on the measuring angle. The beam was focused to the centre of a  reaction  chamber, which allowed measurements at 17\degr, 35\degr, 53\degr, 71\degr, 89\degr, and 107\degr.

\begin{figure}
\begin{center}
\includegraphics[width=0.45\textwidth]{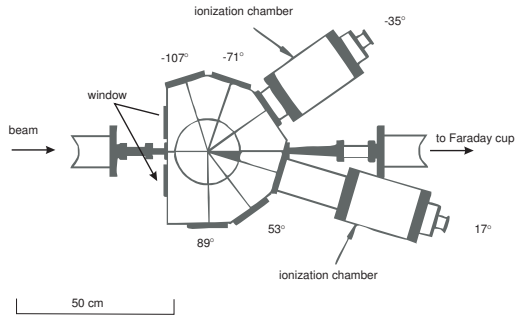}
\caption{Cross section of the reaction chamber.}
\label{Fig:Scattering}
\end{center}
\end{figure}

The layout of this chamber is shown in fig. \ref{Fig:Scattering}.
A target ladder in the centre of the  chamber carried self-supporting  foils    of $^{51}$V (10.0 mg/cm$^{2}$), $^{109}$Ag (10.0 mg/cm$^{2}$), and $^{238}$U (2.9 mg/cm$^{2}$) in addition to a zinc-sulfite foil serving as beam viewer. The beam was focused onto the viewer in the centre of the scattering chamber and then dumped 4 m downstream with the help of a pair of quadrupole magnets into an air cooled Faraday cup. Electrons released in the beam dump were pushed back by an aperture connected to a high voltage. Reaction products were detected with two $\Delta E-E$-telescopes. Each $\Delta E$-detector was a $5\times 5$ cm$^{2}$ Si-diode of 300\, $\mu$m thickness (from micron semiconductor ltd.).  As $E$-detectors we used 4 mm thick Si(Li)-diodes of 6 cm diameter, fabricated in the IKP detector laboratory. The solid angle was defined by  rectangular  apertures made  of  brass yielding $\Delta \Omega = 17.7$ msr and opening angles of 7.6\degr. At the 17\degr position a slightly different geometry resulted into $\Delta \Omega =9.2$ msr and 5.5\degr opening angle. The diodes were followed by standard electronics: commercial charge sensitive pre-amplifiers modified to 4 mV/MeV for the first and 1.9 mV/MeV for the second counters in the experimental hall and main amplifiers and voltage dependent ADC's in an electronic room. The FAZIA collaboration \cite{Carboni12}, \cite{Barlini13} had  recently obtained for a somewhat similar telescope isotopic resolution for elements even up to Z=20 with dedicated electronics and pulse shape discriminators. In the present experiment both detectors operated  behind a window (8 $\mu$m aluminium coated Mylar) and $\approx$ 10 cm air. These spaces were foreseen to host in a later stage ionisation chambers. The isotopic separation was therefore limited only up to beryllium. Spectra for fragments up to sulfur were recorded. The CHIMERA forward detector system consists of $\Delta$E silicon detectors of similar thickness followed by CsI(Tl) scintillators.  Isotopic separation up to oxygen was achieved with these devices \cite{Leneindre02}. Although the resolution of silicon detectors is known to be superior to scintillators, the present set up could not reach this quality.

Finally count rates were transformed to cross sections. Energy losses in the target were taken into account. The energy  calibration with a calibrated precision pulse generator is estimated to be correct to $1.3\%$.  Dead  time  of  the  data  acquisition system was measured and  kept  down  below a $5\%$  level.  The  target thickness is known within $5\%$ error and the number of incident particles to $2\%$. The solid angle was determined with $1.7\%$ error. This leads to a systematic error of $6\%$ in the cross sections.

\section{ Intermediate mass fragment production}\label{sec:Interme-Mass-Fragmen-Product}

\subsection{Double differential cross Section}\label{sub:Double-Differe-Cross-Section}

\subsubsection{Fragments up to oxygen}\label{sub:Fragmen-oxygen}

\begin{figure}
\begin{center}
\includegraphics[width=0.45\textwidth]{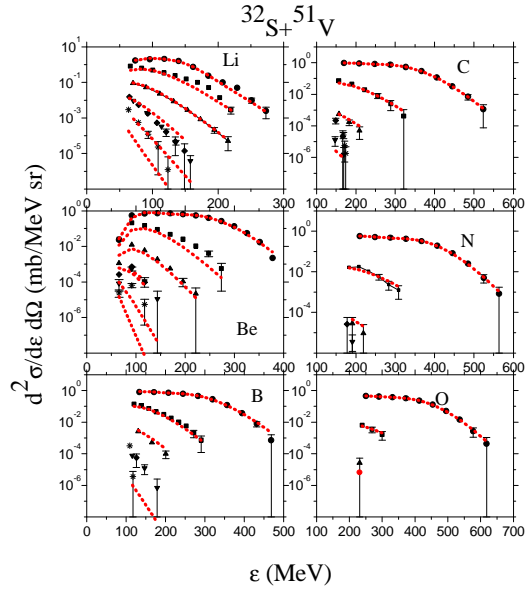}
\caption{(Colour online) Double differential cross sections for
the indicated IMF`$s$ from the interaction of sulfur with vanadium.  The data are shown for laboratory angles 17\degr  (dots),  35\degr
(squares), 53\degr (triangles up), 71\degr (diamonds),
89\degr(triangles down), and 107\degr (asterisks) degrees,
respectively. The error bars indicate the statistical error.  The broken curves are fits with the moving source (MS) model (see section
\ref{sub:Moving-source-analysis}).}
\label{Fig:xvanadium}
\end{center}
\end{figure}

\begin{figure}
\begin{center}
\includegraphics[width=0.45\textwidth]{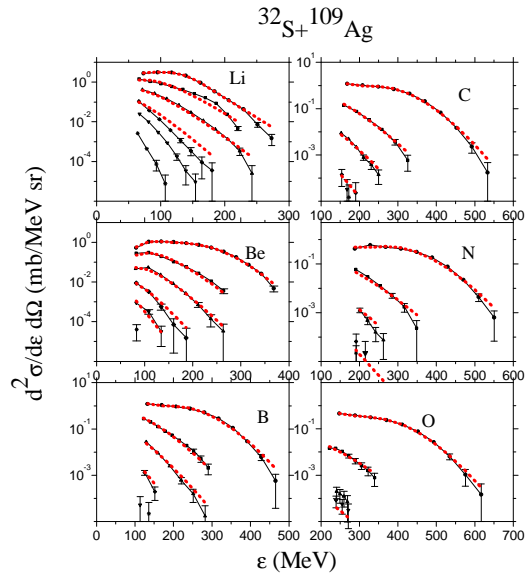}
\caption{(Colour online) Same as fig. \ref{Fig:xvanadium} but for the silver target.}
\label{Fig:xsilver}
\end{center}
\end{figure}

\begin{figure}
\begin{center}
\includegraphics[width=0.45\textwidth]{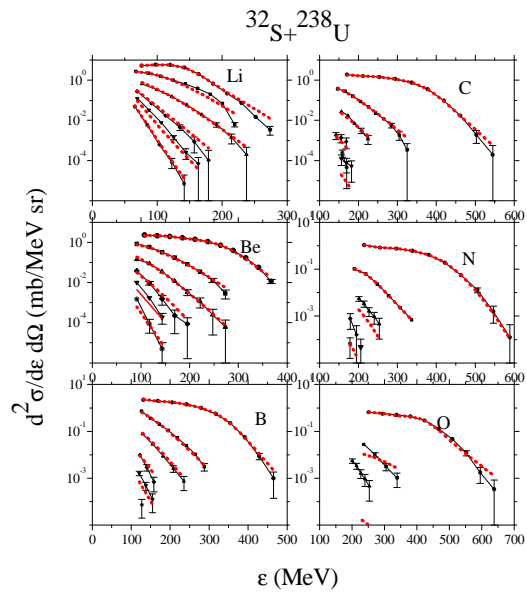}
\caption{(Colour online) Same as fig. \ref{Fig:xvanadium} but for the uranium target.}
\label{Fig:xuranium}
\end{center}
\end{figure}

Figs.~\ref{Fig:xvanadium},~\ref{Fig:xsilver}, and~\ref{Fig:xuranium} show the double differential
cross section for the three targets $^{51}$V, $^{109}$Ag, and $^{238}$U, respectively.

The data are shown for fragments with charge numbers
ranging from 3 to 8. The cross sections follow smooth curves. They strongly decrease with increasing angle and increasing fragment charge number. The cross sections for the smallest angle show up a weak shoulder. The shoulder seems to be more pronounced in the case of the heaviest target. Unfortunately, the
experimental spectra did not cover the full energy range
allowed by kinematics. Low energy particles were not
registered because of finite low-energy detection thresholds
of the telescopes. This lack of information on the low
energy part of spectra could strongly influence the value
of the energy integrated cross section since the spectra
have Maxwellian shape with maxima lying in the
neighbourhood of the energy threshold.

In previous studies performed
by our group a much more pronounced structure was observed and
attributed to stem from a projectile-like system ~\cite{Buh92, Mac85a}.
However, in these systems the most forward measuring angle
was much closer or even below the grazing angle than is the case
for the present study. Such comparisons are also
given for systems similar to the present ones ~\cite{Kim92, Mil90, Wad92, Biz86}. In a study of the system $^{36}$Ar + $^{197}$Au at $E=35$~MeV~A Kim et al.~\cite{Kim92} found a shoulder at the same position with similar strength as in the present study. In $^{40}$Ar + $^{197}$Au reactions at 30 MeV A studied by Milkau et al.~\cite{Mil90} only at 15\degr a weak projectile like component was seen. The authors claim that they could successfully describe the data by fitting only a target-like and an intermediate velocity source. We will come back to this point further down.

In studies of similar systems including angles  below  the  grazing
angle rather strong projectile-like components were identified. Wada  et al.~\cite{Wad92} and Bizard et al.~\cite{Biz86} had  investigated  IMF emission from the $S+Ag$ and $Ar+Ag$ systems at 30 MeV A and 35 MeV A, respectively. In these measurements  a strong component  was observed which corresponds  $\approx 65 \%$ to $80 \%$ of the beam energy/nucleon and hence corresponds to the presently discussed shoulder. However, in these studies a coincidence with a second ejectile was required, so one can not compare the findings directly to the present results. The only data for a reaction studied in this work are due to Lleres et al.~\cite{Lleres87}. They measured energy spectra for the systems
\begin{figure}
\begin{center}
\includegraphics[width=0.45\textwidth]{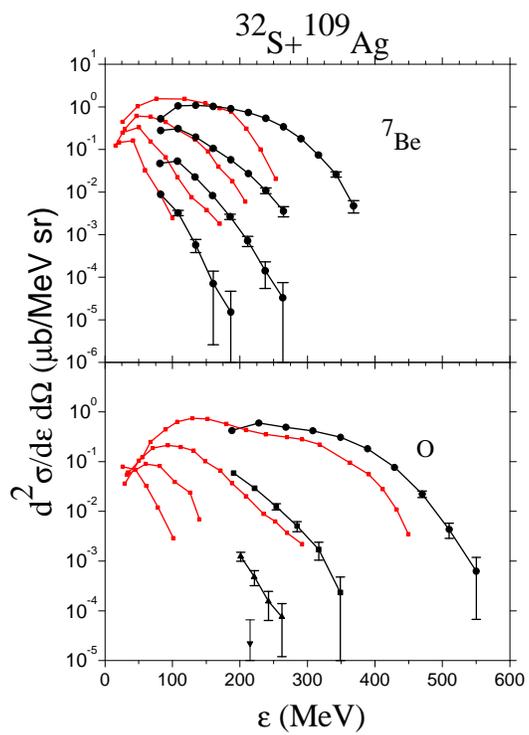}
\caption{(Colour online) Comparison of some present data (solid dots with error bars) with those of Lleres et al.~\cite{Lleres87} at angles of 20\degr, 40\degr, 60\degr and 80\degr .}
\label{Fig:Lleres}
\end{center}
\end{figure}
sulfur on natural silver and on gold  at 30 MeV/nucleon. However, their detector consisted of much thinner counters than those in the present experiment. Therefore the data cover mainly the low energy part which was not accessible in the present experiment. On the other hand fragments with very high energies were not detected. Some data are shown together with the present one in fig. \ref{Fig:Lleres}. They compare favourable with each other although the emission angles are not identical.

\subsubsection{Spectral shapes}\label{sub:Spectra-shapes}

\begin{figure}[h]
\begin{center}
\includegraphics[height=0.3\textwidth]{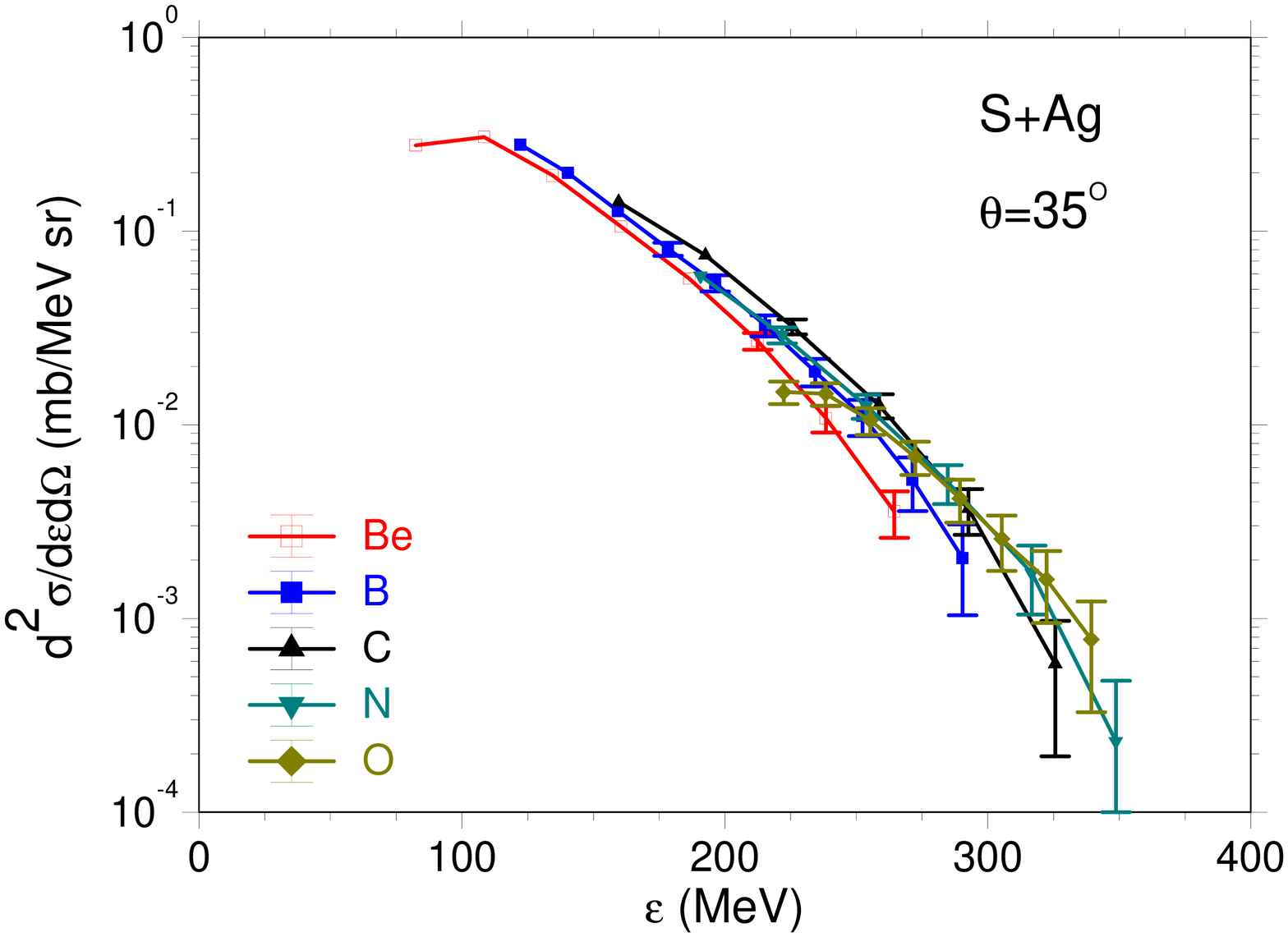}
\caption{(Colour online) The double differential cross section for the reaction $^{32}$S+$^{109}$Ag at $\theta=35$\degr. The data are shown by different symbols with error bars indicating the different fragments. The curves are just connections between the data points. }
\label{Fig:Spec_V}
\end{center}
\end{figure}

Inspection of the spectral shapes in Figs. \ref{Fig:xvanadium}-\ref{Fig:xuranium} shows for the angles larger than 17\degr an almost exponential slope with a slight curvature. This is valid of course only for the energy range covered by the present experiment. The slope parameter decreases with increasing angle, because the emission is forward peaked. We have then plotted the spectra for all ejectile types at the same angle and the same target into one coordinate system. By way of example we show such a plot for the silver target and an angle of 35\degr in fig. \ref{Fig:Spec_V}.
Surprisingly the yields agree with each other within $\pm 20\%$. A closer look shows that the smallest cross section is for beryllium which are even-odd  isotopes. The most probably odd-odd nuclei (boron and nitrogen) have smaller cross sections than the even-even nuclei (carbon and oxygen). Such a pairing effect was studied in more detail in Refs. \cite{Piantelli13}, \cite{Agostino12} and \cite{Lombardo11}. We will come back to this point in subsection \ref{sub:Elemental_Yields}.

\subsubsection{ Fragments up to sulfur}\label{sub:Fragmen-sulfur}

\begin{figure}
\begin{center}
\includegraphics[width=0.450\textwidth]{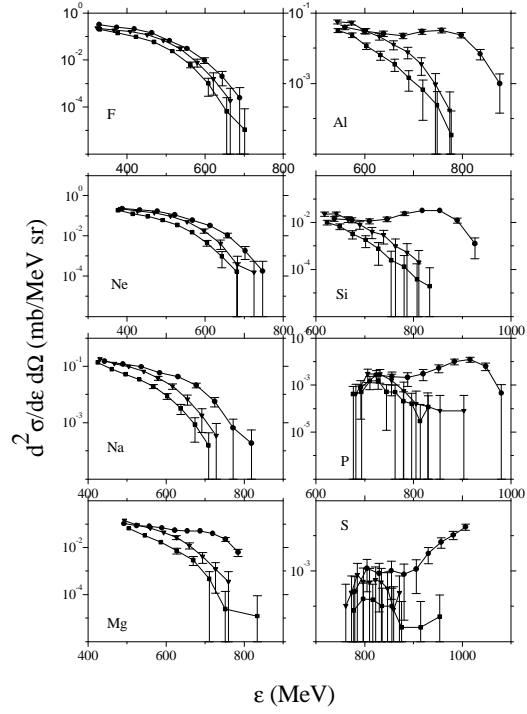}
\caption{IMF cross sections  for  the  three target materials U (dots), Ag (triangle down) and V (square) measured at a laboratory angle of 17\degr. }
\label{Fig:IMF_Z}
\end{center}
\end{figure}

In the following we will concentrate on heavier IMF's up to sulfur.
Such IMF's could only be measured at 17\degr.  In  fig.~\ref{Fig:IMF_Z}  the
double differential cross sections for the targets $^{51}$V, $^{109}$Ag, and $^{238}$U are shown.

\begin{figure}
\begin{center}
\includegraphics[width=0.450\textwidth]{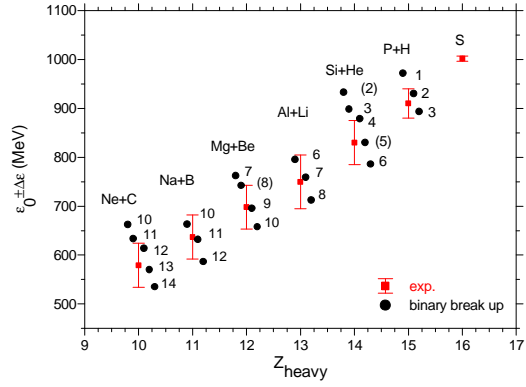}
\caption{(Colour online) The centroids of Gaussians fitted to high energy component are shown as dots and the bars denote their widths. Also shown are the positions for binary elastic break up calculated with eq. \eqref{eq:two-body}) }
\label{Fig:Break_up}
\end{center}
\end{figure}
The lowest energy measured is given by the thickness of the $\Delta E$ detector.  The yields decrease with increasing mass number as it was the case for the lighter fragments. In addition, the largest  energy per charge decreases with increasing mass for the two lighter target nuclei. A value of 60 MeV per charge, which is very close to the projectile velocity, is the largest energy for the heaviest fragments but is not for the lighter ones. The situation for the uranium target is somewhat different. Although the cross sections for  the lighter fragments are similar to those for the other targets, for the heavier fragments a new component emerges around 60 MeV/charge.  This becomes clearly visible for fragments heavier than neon and dominates the double differential cross sections for very heavy fragments. It seems that in the  case of the uranium target, different projectile-like components exist. While the previously discussed component  around 45~MeV/charge decreases with increasing mass, the additional component increases relatively to the total yield. Its centroid depends only weakly on the fragment type. Similar shapes are found in the interaction of oxygen with gold at the same beam energy per nucleon~\cite{Rama08}.  Because $\varepsilon /Z$ is almost the  velocity  squared, we can state that only one source is responsible for this component. To account for this  source in a more quantitative way Gaussian distributions were fitted to this component.

The obtained centroids $\varepsilon_0$ and widths $\Delta\varepsilon$ are
shown in fig.~\ref{Fig:Break_up}. The widths as function of
$\varepsilon/Z_f$ are almost independent of the fragment masses. The
overall feature is that of a quasi elastic process. We may estimate
the centroid of the energy distribution within the peak. Suppose the
process is
\begin{equation}\label{eq:binary}
A_{proj}\to A_1+A_2.
\end{equation}
We then expect the centroid to be
\begin{equation}\label{eq:two-body}
\varepsilon_0 =E_{proj}\frac{A_1}{A_{proj}}+Q
\end{equation}
with $Q$ the $Q$-value of the process (\ref{eq:binary}). The
corresponding values are indicated in fig. \ref{Fig:Break_up} as
points next to the fit result for $Z_1$ for different $A_2$.

This component is only observed in case of the uranium target, where
the detector opening just covers the grazing angle. It may well be
that this type of break up is mainly in the Coulomb field of the
target nucleus.

It should be mentioned that for carbon,  nitrogen  and  oxygen   the
largest fragment velocities were observed which extend up to  twice  the
beam velocity. This is true for  all   targets  studied  (see  also
figs.~\ref{Fig:xvanadium},~\ref{Fig:xsilver}, and~\ref{Fig:xuranium}).  However,  in  terms   of   transferred
momentum there is a continuous increase with  decreasing  fragment  mass
for the highest velocity components.

\subsection{Isotopic intensities}\label{sub:Isotopic}

\begin{figure}
\begin{center}
\includegraphics[width=0.45\textwidth]{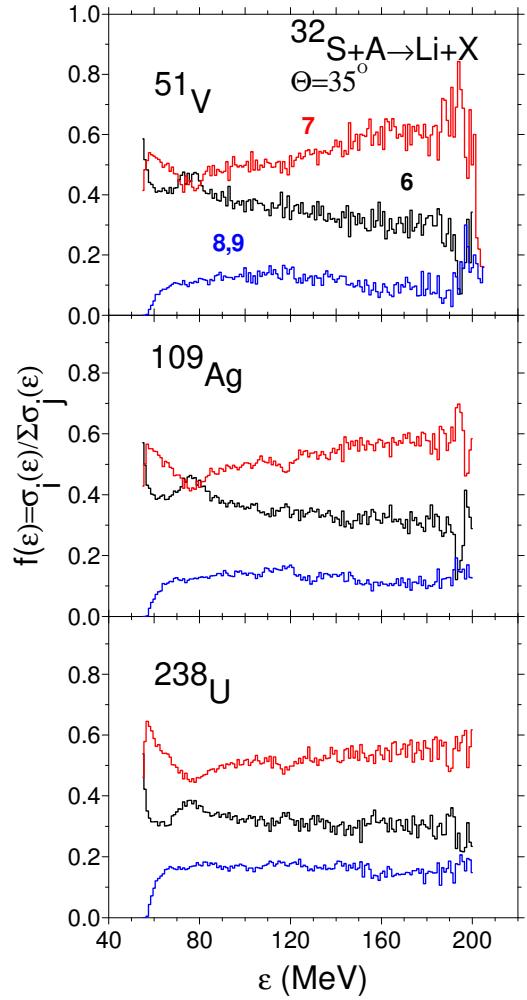}
\caption{(Colour online) The ratios $d^2/d\sigma(^jLi)/d^2\sigma(Li)$ with $j=6, 7, 8$ for the different target nuclei as function of the lithium fragment energies.}
\label{Fig:LI_ratio}
\end{center}
\end{figure}

So far only elemental yields were discussed. Another quantity  rarely  studied  in	the  literature  is  the energy dependence of isotope ratios. In the  present  experiments  such  ratios could only be determined for lithium and beryllium isotopes. In fig.~\ref{Fig:LI_ratio} the energy dependencies of the ratios of lithium isotopes to the total lithium emission at a laboratory angle of 35 degrees are shown.

They look not very different for all three targets. The largest probability to be emitted has $^{7}Li$, which is the isotope closest to the stability line. The next probable isotope is $^{6}$Li. All energy dependencies are smooth. It is interesting to notice that around $\varepsilon \cong 75$ MeV  the  emission of  stable isotopes is almost the same. The isotopic ratios are almost independent of  the target mass although the ratio $(N_{P}+ N_{T})/(Z_{P}+Z_{T})$  varies  from 1.5  for uranium to 1.13 for vanadium. The ratio $\sigma (^{7}Li)/\sigma (^{6}Li)$ was found to  vary in that range by one order of magnitude~\cite{Wad87,Dea91}.  The  reason for this constancy is unclear at the moment.

\begin{figure}
\begin{center}
\includegraphics[width=0.45\textwidth]{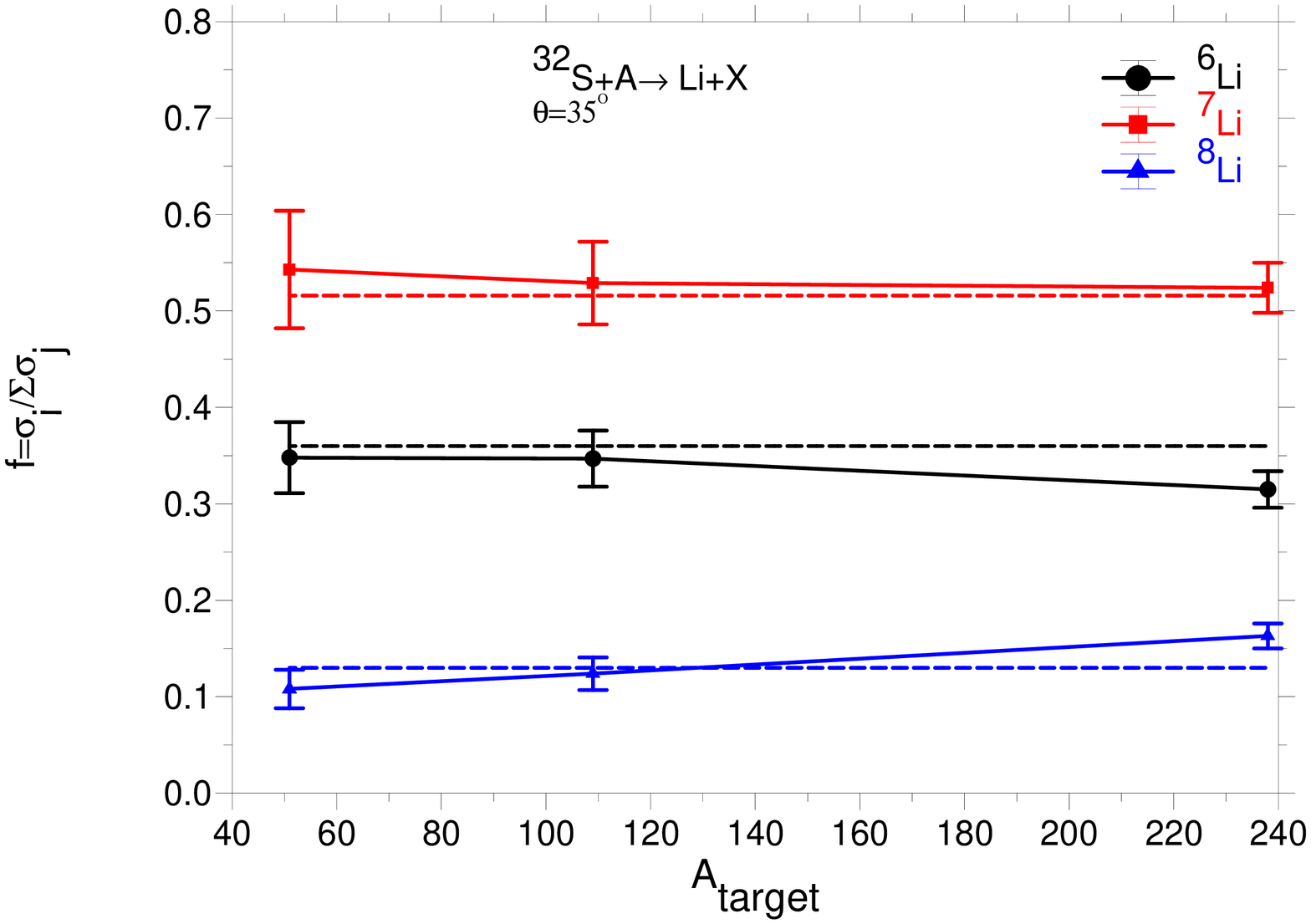}
\caption{(Colour online) Energy integrated ratios from fig. \ref{Fig:LI_ratio}. The solid line connects the data points, the dashed lines are arithmetic mean values.}
\label{Fig:Li_integ}
\end{center}
\end{figure}

We proceed and study the energy integrated ratios.  In fig.~\ref{Fig:Li_integ} these ratios are shown as function of the target mass. As expected from the double differential ratios there is only a very weak mass dependence for the differential ratios.
Careful inspection gives that the fractions of $^{6}$Li emission drops with increasing target mass while those for $^8$Li increases. This is a clear indication that an increase of the neutron fraction in the emitting system increases the fraction of the isotope with higher neutron excess. Such an effect was also seen by the CHIMERA collaboration \cite{Filippo14}. For the neutron deficient system $^{78}$Kr+$^{40}$Ca the strongest carbon isotope was $^{12}$C followed by $^{13}$C and $^{11}$C. For the more neutron rich system $^{86}$Kr+$^{48}$Ca the most frequent isotope was $^{13}$C followed by $^{12}$C and $^{14}$C.

\begin{figure}
\begin{center}
\includegraphics[width=0.45\textwidth]{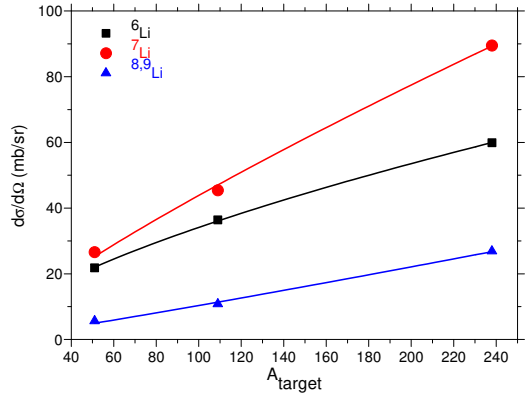}
\caption{(Colour online) Energy integrated cross section for the emission of lithium isotopes at a laboratory angle of $\theta$=35\degr as function of the target mass number. The different isotopes are indicated in the figure by different symbols. The solid curves are fits with an $A_\text{target}^{\alpha}$ law.}
\label{Fig:Comp_Li}
\end{center}
\end{figure}

It is interesting to study the target mass dependence directly. Therefore, we do not look at relative yields but to absolute yields. For this purpose we integrate energy spectra for the different isotopes and get differential cross sections. For lithium isotope emission at an angle of 35\degr they are shown in fig. \ref{Fig:Comp_Li}. The cross sections follow a power law $A^{\alpha}$, which is also indicated in the figure. The parameter $\alpha$ is 2/3 for $^6$Li and 1 for $^{8,9}$Li. There is a sequence $\sigma(^7\text{Li})>\sigma(^6\text{Li})>\sigma(^{8,9}\text{Li})$ which is independent of the target nucleus.

\begin{figure*}[htb]
\begin{center}
\includegraphics[height=0.4\textheight]{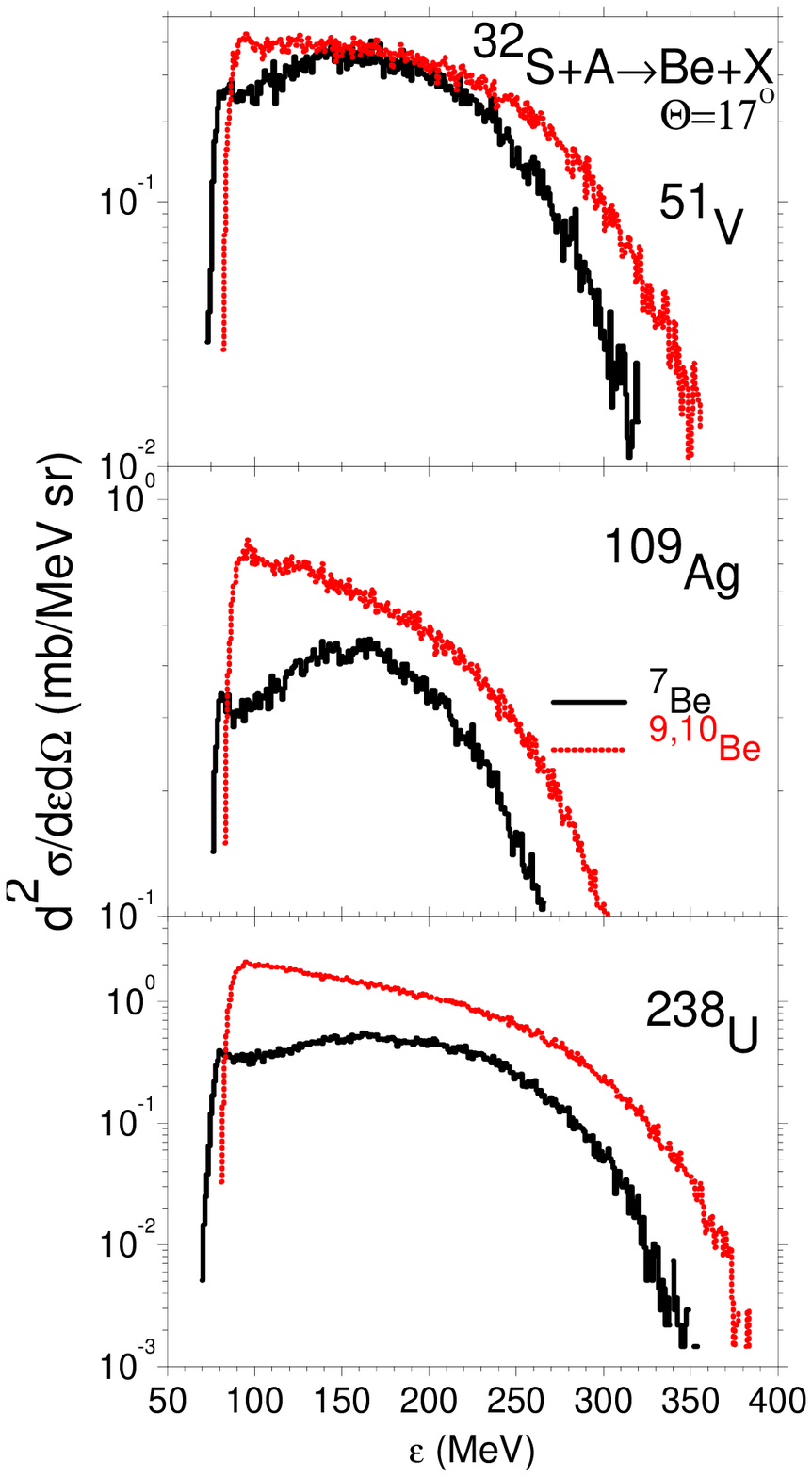}
\includegraphics[height=0.4\textheight]{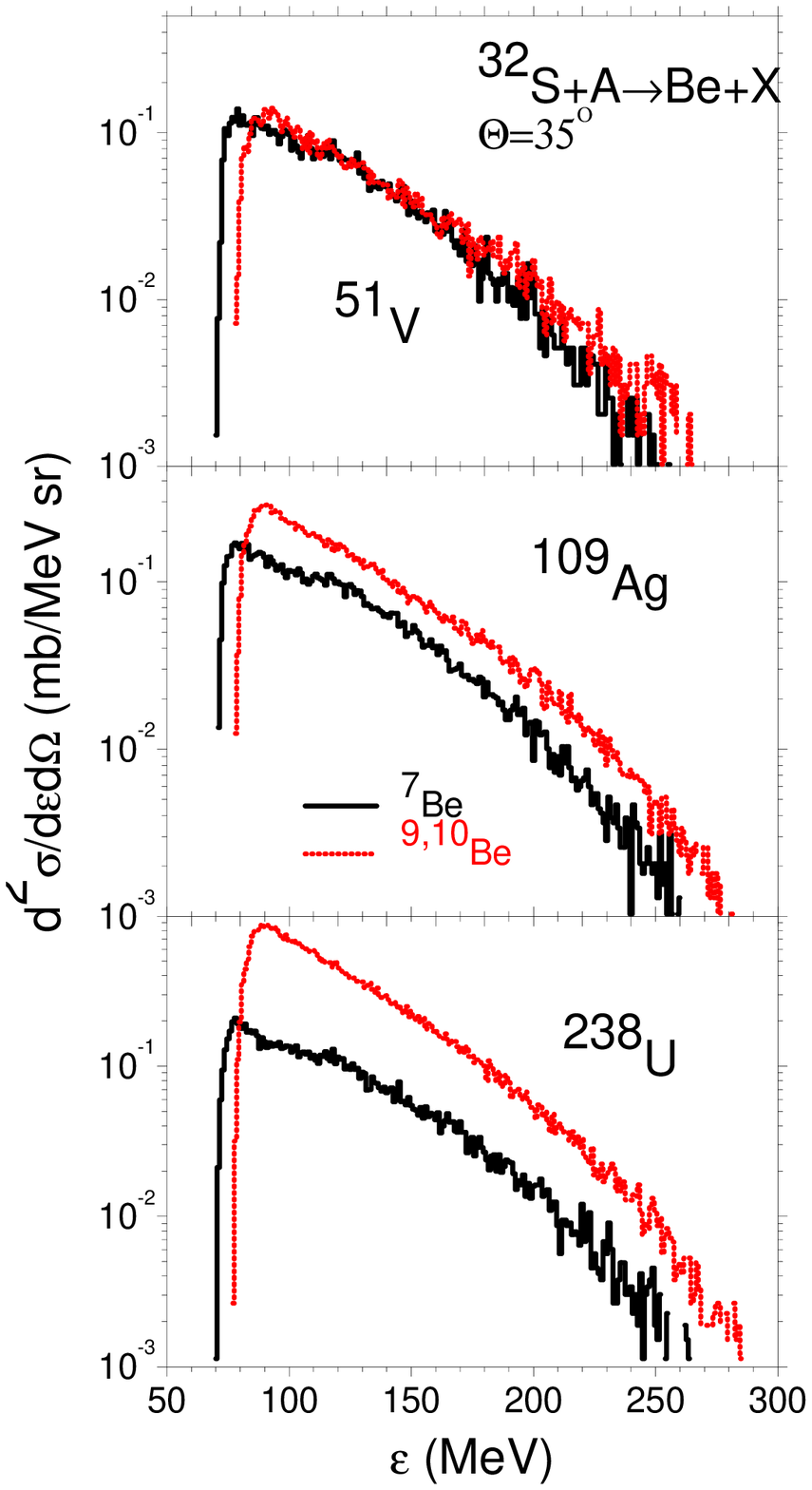}
\includegraphics[height=0.4\textheight]{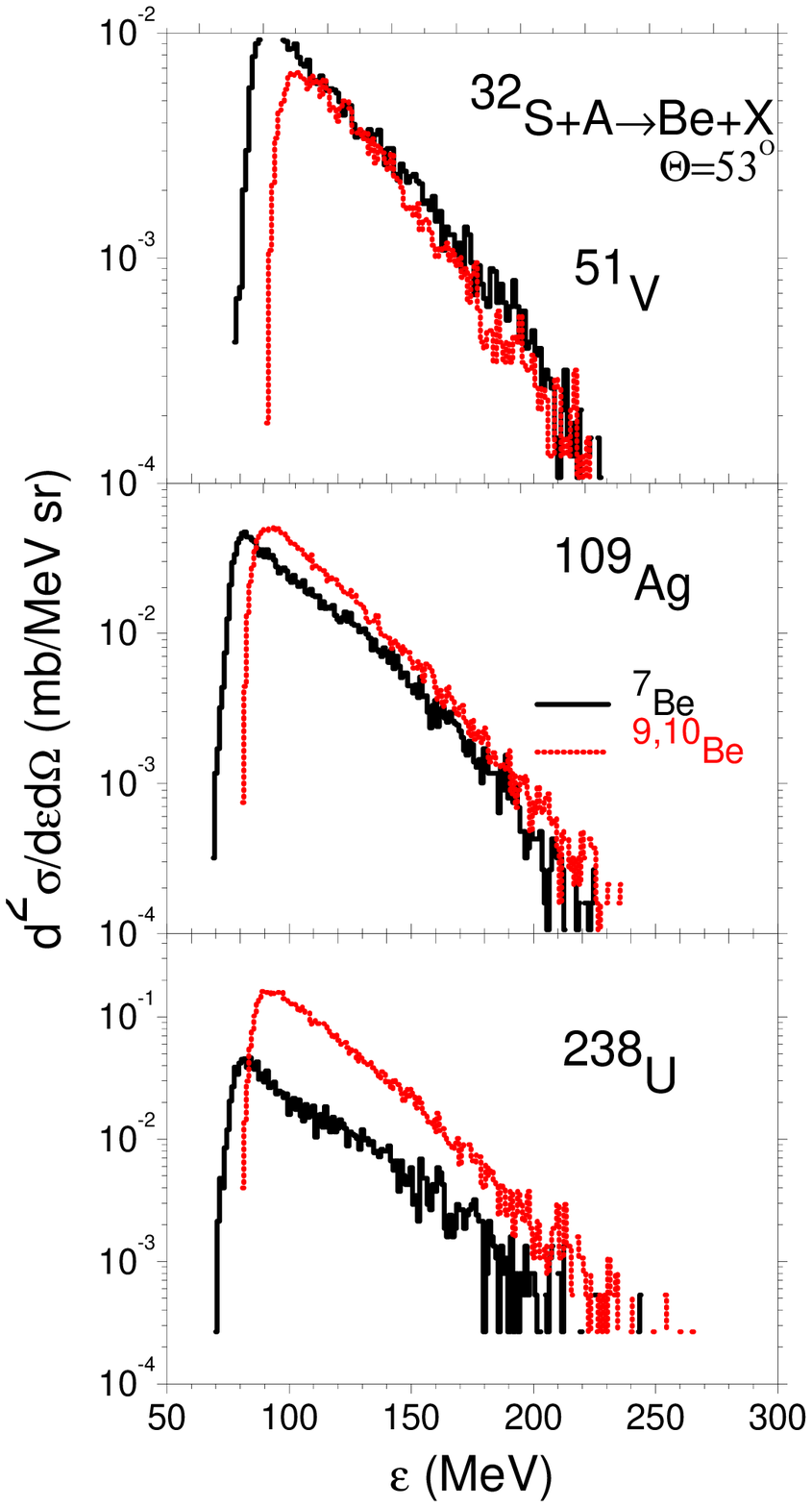}
\caption{(Colour online) The double differential cross sections for the indicated reactions.}
\label{Fig:Be7-10}
\end{center}
\end{figure*}

The situation is quite different in the case of beryllium isotopes. In this case we could only distinguish $^{7}$Be and $^{\geq 9}$Be  cross sections, where the latter include those from heavier isotopes. fig.~\ref{Fig:Be7-10} shows the  energy  spectra for the two groups and for the angles 17\degr, 35\degr and 53\degr. While for the vanadium target the cross sections for the two groups are almost the same for all angles there is a tendency that $\sigma(^7\text{Be})<\sigma(^{\geq 9}\text{Be})$, most evident for the uranium target. The spectral shape at 17\degr is bell shaped while for the larger angles it looks like exponentials. This feature was also found for the elemental cross sections (see section \ref{sub:Spectra-shapes}). The largest difference between the two isotope groups is for the uranium target at 53\degr. The cross section for $^7\text{Be}$ is in the order of 15$\%$ to 25$\%$ of the total Be cross section whereas for the other cases it ia around $50\%$.

While for the different isotopes of one element the Cou\-lomb energy of the emitting systems is almost the same, possible Coulomb  effects can be studied by the ratio of isotopes from different elements but same mass number. By way of example we study the ratio
\begin{equation}
R = \frac{d\sigma (^7\text{Be})}{ d\sigma (^7\text{Li})}
\label{EQN_9}
\end{equation}
as measured at 35\degr.  For that  purpose we introduce
a reduced Coulomb barrier$\tilde V_C$
\begin{equation}
\tilde V_{C}=V_{C}/Z_{f}
\label{EQN_10}
\end{equation}
which only very weakly depends on the fragment charge
$Z_{f}$ for isobaric isotopes. We find a perfect linear dependence of the ratio on the reduced Coulomb barrier
\begin{equation}
\frac{d\sigma (^{7}\hbox{Be})}{ d\sigma (^{7}Li)} = 0.407(3) -
0.0188(2)\tilde V_{C}
\label{EQN_11}
\end{equation}
giving for vanishing Coulomb effects a ratio of 0.4  which could
\begin{figure}
\begin{center}
\includegraphics[width=0.45\textwidth]{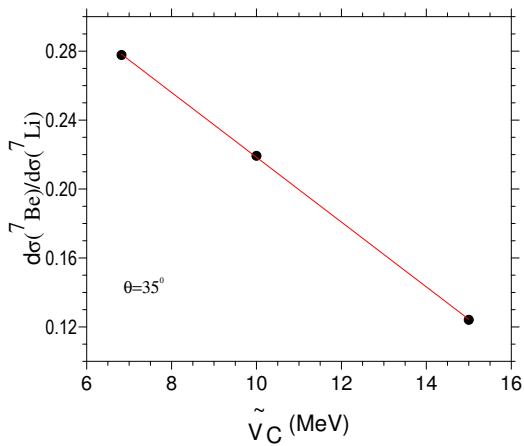}
\caption{The ratio of the cross sections for the two isobaric isotopes $^7$Be and $^7$Li as function of the reduced Coulomb barrier $\tilde V_{C}$. The solid line is the dependence eq. \eqref{EQN_11}}.
\label{Fig:R_Li_Be}
\end{center}
\end{figure}
perhaps be explained  by  combinatorial  effects~\cite{Boa84}. However,
more data points are needed to fully establish such a relationship  as
eq. \eqref{EQN_11} or the mass dependencies of the slope parameters.

From the above  discussion we conclude  that  it  is  impossible  at present to decide whether the different slopes of  the elemental  yields are due to different available kinetic  energies  or  due  to mass  or Coulomb effects. The same ratio $R$ was also studied by Lombardo et al. \cite{Lombardo10} in the reactions $^{40}$Ca+$^{40}$Ca, $^{40}$Ca+$^{46}$Ti, and $^{48}$Ca+$^{48}$Ca. Similar as here the ratio $R$ is largest for the neutron deficient system. This finding holds for a fitted mid-velocity component as well as for a projectile like component. These results are in accord with scenarios that link this effect to isospin drift and diffusion processes in the nuclear medium \cite{Tsang09}.

\section{Model analysis}\label{sec:Model-analysis}

\subsection{Moving source analysis}\label{sub:Moving-source-analysis}

As was discussed above such data as the present ones are often analysed in terms of moving sources. Here we will give only a very short description, more details are found elsewhere~\cite{Mac85a}. Isotropic Maxwell--Boltzmann distributions in the rest frame of the emitting systems lead to
\begin{equation}
\frac{d^2\sigma(\varepsilon,\theta)}{d\varepsilon d\Omega} =\\
\sum_{i=1}^nc_i\sqrt{\varepsilon} \exp\left[\frac{-\left(\varepsilon-\sqrt{2m\varepsilon} v_i\cos\theta +mv_i^2/2\right)}{T_i} \right].
\end{equation}
With $v_i$ the velocity of the i-th source is denoted and by $T_i$ its temperature (measured in MeV). The angle and energy integrated cross section for the component $i$ is given by
\begin{equation}
\sigma_i = 2c_i(\pi T_i)^{3/2}.
\end{equation}
In proton induced reactions the PISA collaboration \cite{Bubak07}, \cite{Budzanowski10} found two sources sufficient to account for the data: one source is an equilibrated system (compound nucleus) and the second one an intermediate source which emits particles  more forward into the beam direction than the compound nucleus. Here we should in principle consider one more source: a projectile like source. However, in the PISA experiment data could be measured at rather small energies down to the Coulomb barrier. By way of example this is approximately 15 MeV for $^6$Li emission from $p$+Ni interactions. Here for the rather similar case the lowest  energy detected is 100 MeV. Therefore the present data are not sensitive to compound nucleus emission and we have fitted only a projectile like source and an intermediate source.

In experiments employing proton beams barrier penetration was taken into account  \cite{Bubak07}, \cite{Budzanowski10}, \cite{Mac06}. Here we treat the influence of the Coulomb force by a shift $\varepsilon=\varepsilon'+V_\text{Coul}$ with $\varepsilon'$ the energy of the fragment before boosted by the Coulomb repulsion.

We have fitted the quantities $\sigma_i$, $v_i$ $T_i$ and one Coulomb energy $V_\text{Coul}$ to the
double differential cross sections for the IMF's from lithium to
oxygen. The fits are shown in Figs \ref{Fig:xvanadium}, \ref{Fig:xsilver}
and \ref{Fig:xuranium}. Although the $\chi^2$-values achieved
indicate poor fits the data were accounted for. The velocities of the fast projectile like source are typically 0.2$c$. The beam velocity corresponds to $v_\text{beam}=0.26c$ so that $v_\text{proj}/v_\text{beam}\approx 0.77$. There is a general increase of the velocity with increasing charge number. This can be an indication that the heavier fragments are emitted earlier than the light ones. The same finding is true for the intermediate source. In this case the ratio is $v_\text{int}/v_\text{beam}\approx 0.50$. However, the Coulomb force can also the origin of this effect.

 In general the velocities of the fast components are close to the beam velocity, the intermediate component has half the beam velocity and, when a third component is fitted in addition,  this component has velocities of approximately the fully equilibrated system, i.e. complete momentum transfer. There is a general increase of the velocity with increasing charge number.

\subsection{ Generalised moving source analysis}

A variation of the moving source was given in Ref.~\cite{Mach90}. Instead of a finite sum of sources with discrete source velocity and temperature it is assumed that the cross section is given by
\begin{multline}
\frac{d^{2}\sigma (\varepsilon ,\vartheta )}{d\varepsilon  d\Omega} =\\ C
\sqrt{\varepsilon } \exp \left\{ -\left[ \varepsilon
-2\sqrt{\varepsilon E \left(\varepsilon \right)} \cos \left(\vartheta
\right)   +E_{0}\left(\varepsilon
\right)\right] /T\left(\varepsilon \right)\right\}
\label{EQN_2}
\end{multline}
with $E(\varepsilon)=1/2 mv_0^2$ the kinetic energy of the source and the temperature $T(\varepsilon)$, both functions of the ejectile energy $\varepsilon$. This what is called the generalised moving source model (GMSM) \cite{Machner90} \cite{Buh92}. If eq. (\ref{EQN_2}) is the right
description
for the cross section, then the data follow straight lines on a plot
\begin{equation}\label{eq:apar}
\ln  \frac{d^{2}\sigma (\varepsilon ,\vartheta )}{d\varepsilon  d\Omega} = b(\epsilon)+a(\epsilon)\cos (\vartheta )
\end{equation}
for a constant ejectile energy  with the slope parameter being
\begin{equation}
a(\varepsilon )=\frac{2\sqrt{\varepsilon  E(\varepsilon )}} {T(\varepsilon )} .
\label{EQN_4}
\end{equation}
From these equations (\ref{EQN_2}-\ref{EQN_4}) it follows
\begin{equation}
\frac{d^{2}\sigma (\varepsilon ,\vartheta_i )}{d\varepsilon  d\Omega} = \frac{d^{2}\sigma (\varepsilon ,\vartheta_j )} {d\varepsilon  d\Omega} \exp \left\{a\left(\epsilon\right) \left[\cos\left(\vartheta_j\right))-\cos\left(\vartheta_i\right)\right]\right\},
\label{eq:gmsm_angles}
\end{equation}
which implies that when the cross section at one angle is known it is then given at other angles. From this equation it becomes clear that the function $a(\epsilon)$ plays an eminent role within this model. Since it occurs within an exponent its certainty is crucial for the predictions of the model.

The eminent role of this parameter becomes further evident when looking to the angle integrated cross section, which is within this model \cite{Mac06}
\begin{eqnarray}
\frac{d\sigma(\epsilon)}{d\epsilon} &=& \frac{2\pi}{a}(e^{b+a}-e^{b-a})\\
&=& \frac{4\pi}{a(\epsilon)}\frac{d^{2}\sigma (\varepsilon ,\vartheta )} {d\varepsilon  d\Omega} \frac{\sinh[a(\epsilon)]}{\exp [a(\epsilon) \cos(\vartheta)]}.
\end{eqnarray}
Again the cross section at one angle together with the energy dependence of the parameter $a$ determines the angle integrated cross section.

\begin{figure}[h]
\begin{center}
\includegraphics[width=0.45\textwidth]{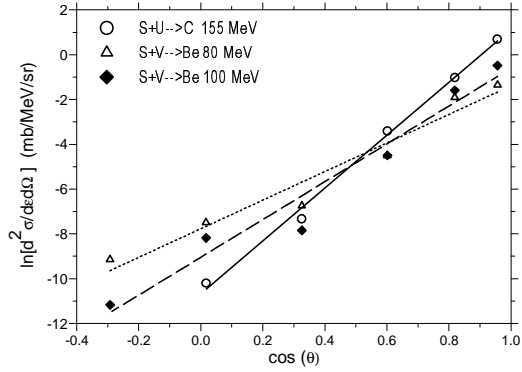}
\caption{Fit of eq. (\ref{eq:apar}) to some selected data.}
\label{Fig:apar-fit}
\end{center}
\end{figure}
In order to extract $a(\epsilon)$ the function eq. (\ref{eq:apar}) was fitted to different angular distributions for different reactions. It is shown in fig. \ref{Fig:apar-fit} for some angular distributions. However, the data have to be excellent with respect to energy range and angles covered. This is demonstrated for the reaction $^{32}$S+$^{238}$U$\to$C+X.
\begin{figure}[h]
\begin{center}
\includegraphics[width=0.45\textwidth]{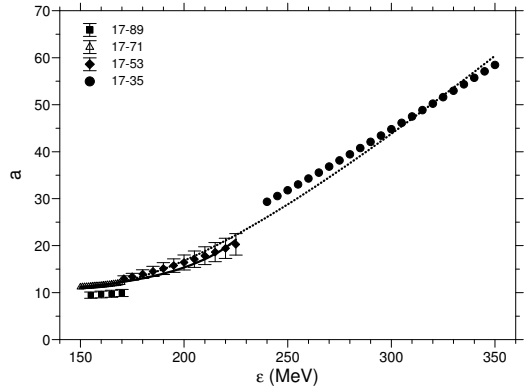}
\caption{The $a(\epsilon)$-parameter deduced from the reaction $^{32}$S+$^{238}$U$\to$C+X. The different angular ranges included in the fits are indicated by different symbols indicated in the figure. Also shown are fits to results at low energies (soli curve) and the total data set (broken curve).}
\label{Fig:apar_uran}
\end{center}
\end{figure}
The results are shown in fig. \ref{Fig:apar_uran}. There are obviously four sets of results: at the smallest energies the small values for four angles ranging from 17\degr to 89\degr. The larger values are from the range 17\degr to 71\degr, which contains four angles. The middle set covers the angular range from 17\degr to 53\degr while the largest values at the highest energy cover the to most forward angles 17\degr and 35\degr. Fits which includes all data points can not account for the results in the energy interval 200 till 300 MeV. This is a result of the inconsistency of the extracted parameters. This inconsistency is also reflected when looking to the angular distributions.
\begin{figure}[h]
\begin{center}
\includegraphics[width=0.45\textwidth]{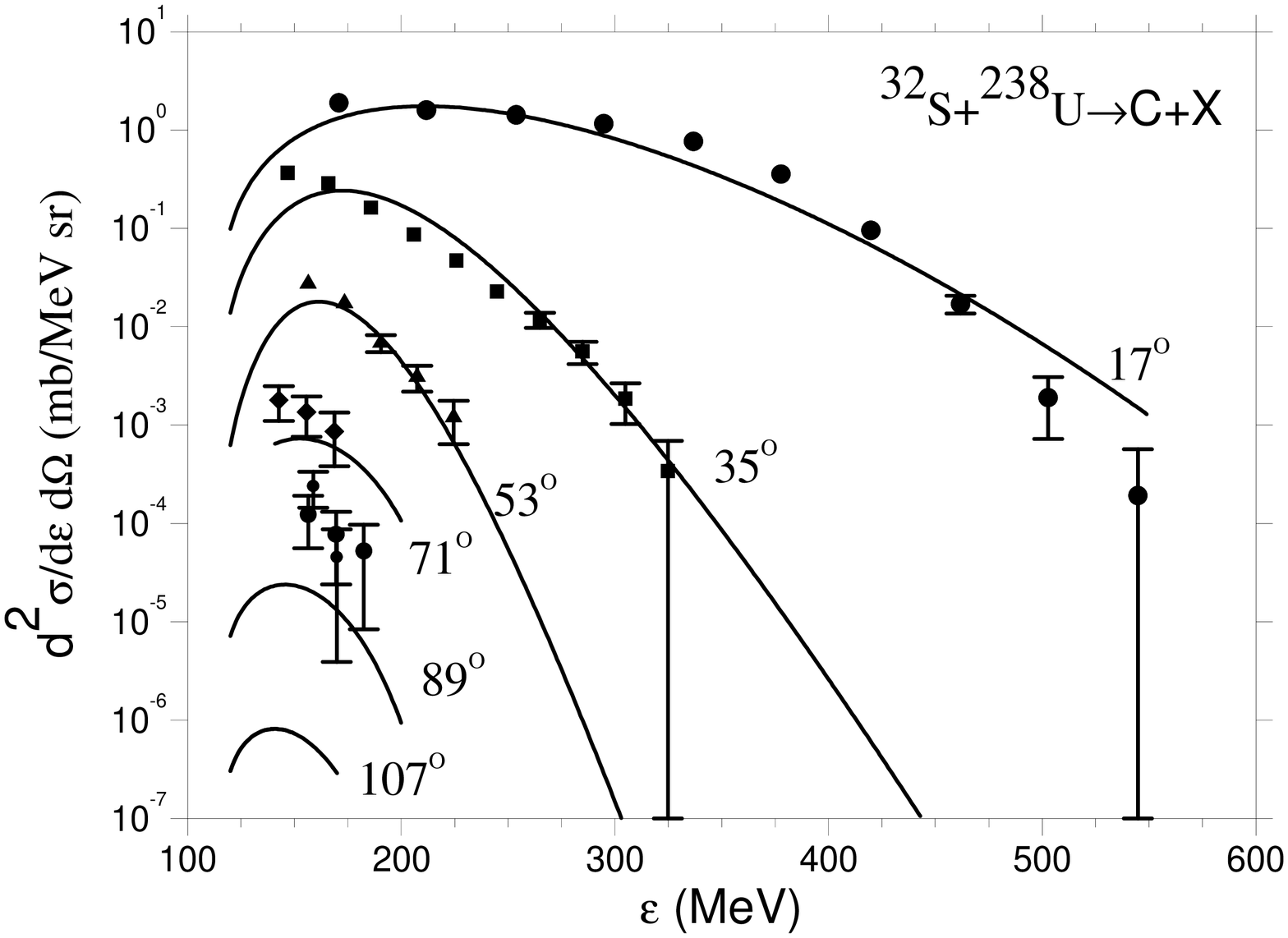}
\caption{Energy spectra of carbon fragments for the angles indicated in the figure. For curves see text.}
\label{Fig:GMSM_U}
\end{center}
\end{figure}
This is plotted in fig. \ref{Fig:GMSM_U}. Shown are the cross sections for the reaction $^{32}$S+$^{238}$U$\to$C+X from fig. \ref{Fig:xuranium}. In a first step the energy dependence of the parameters in eq. \ref{EQN_2}, $c$, $E$ and $T$, where fitted (see further down), in order to reproduce the energy spectrum at 17\degr. Then the spectra at 35\degr and 53\degr were calculated using eq. \ref{eq:gmsm_angles} and the fit to $a(\epsilon)$ ranging up to the highest energies. This method reproduces the cross sections at these two angles. However, it fails for the larger angles. We then applied the low energy solution for $a(\epsilon)$. The model prediction are shown for the three largest angles. The quality of the model predictions is less satisfactory here. At this point we may conclude that GMSM seems to contain the correct physics, but at least the present data do not cover cross sections at a sufficient number of angles and wide enough energy ranges.

\begin{figure}[h]
\begin{center}
\includegraphics[width=0.45\textwidth]{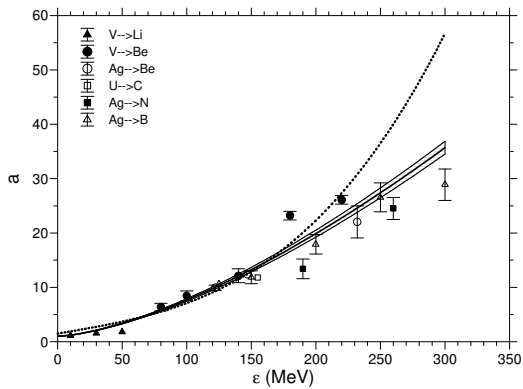}
\caption{The parameter $a(\varepsilon)$ as function of the ejectile
energy. The solid curve is a fit to present data and the shaded area its variance. Also shown is the result ~\cite{Buh92} (dotted curve) obtained for $^{15}$N and $^{20}$Ne induced reactions at 530 MeV.}
\label{Fig:apar}
\end{center}
\end{figure}

We then proceed extract $a(\epsilon)$-parameters for different reactions at different ejectile energies. The results are shown in fig. \ref{Fig:apar}.
To these results a smooth curve was fitted: $a(\epsilon)=1+(0.006677\pm 0.000223)\epsilon^{3/2}$ which is also shown in the figure. It is interesting to note that the present result is undistinguishable for $\epsilon \leq 200$MeV from those obtained for $^{15}$N and $^{20}$Ne induced reactions at approximately the same energy per nucleon \cite{Buh92}. We make use of this solution for $a(\epsilon)$ to extract angle integrated cross section from which total cross section were obtained by numerical integration.

\begin{figure}[h]
\begin{center}
\includegraphics[width=0.45\textwidth]{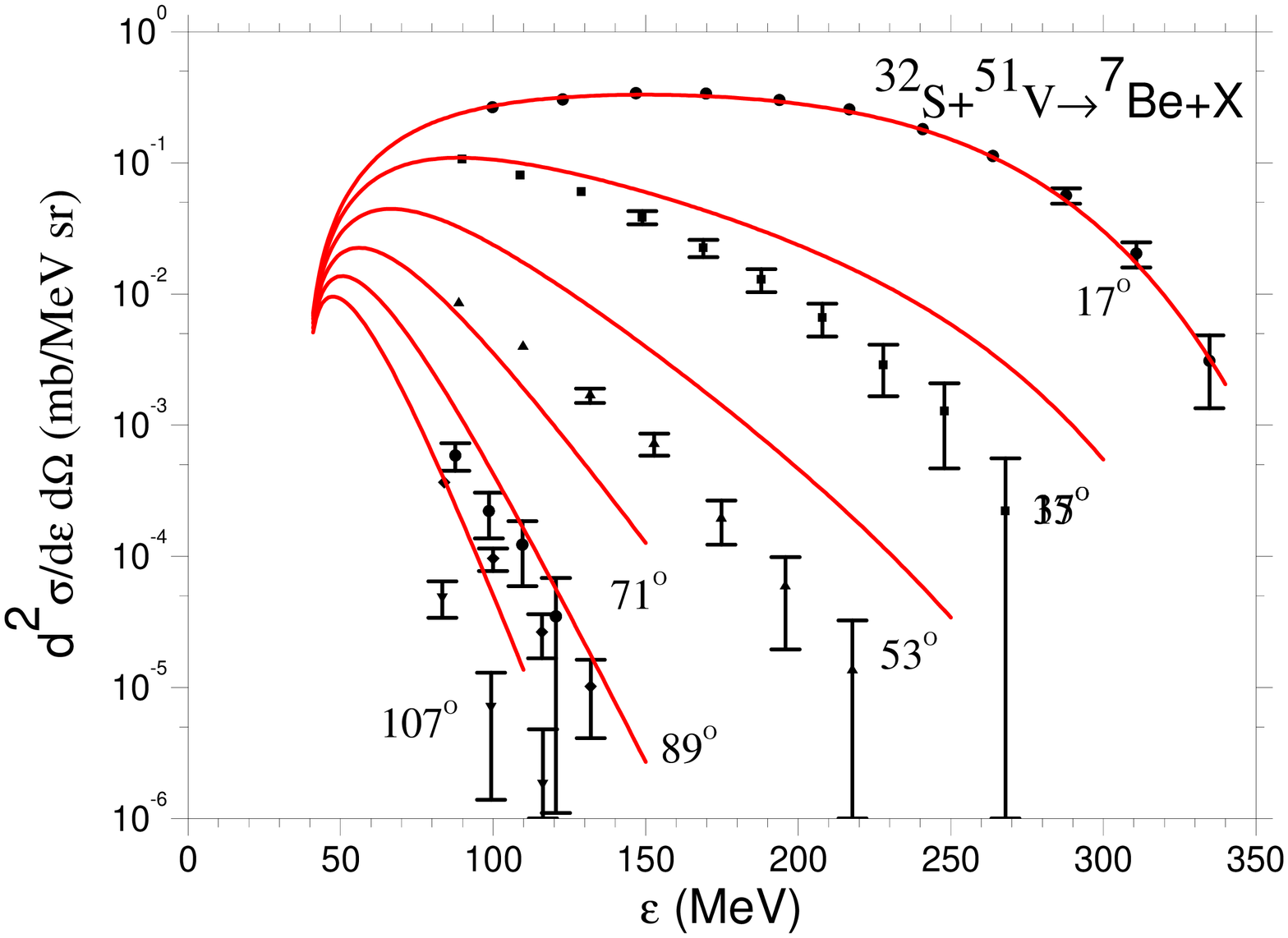}
\caption{(Colour online) Spectra for the $^{51}$V($^{32}$S,$^7$Be)X reaction at the angles indicated next to the appropriate curves. The fit and predictions of the GMSM are shown as solid curves.}
\label{Fig:V}
\end{center}
\end{figure}

\begin{figure}[h]
\begin{center}
\includegraphics[width=0.45\textwidth]{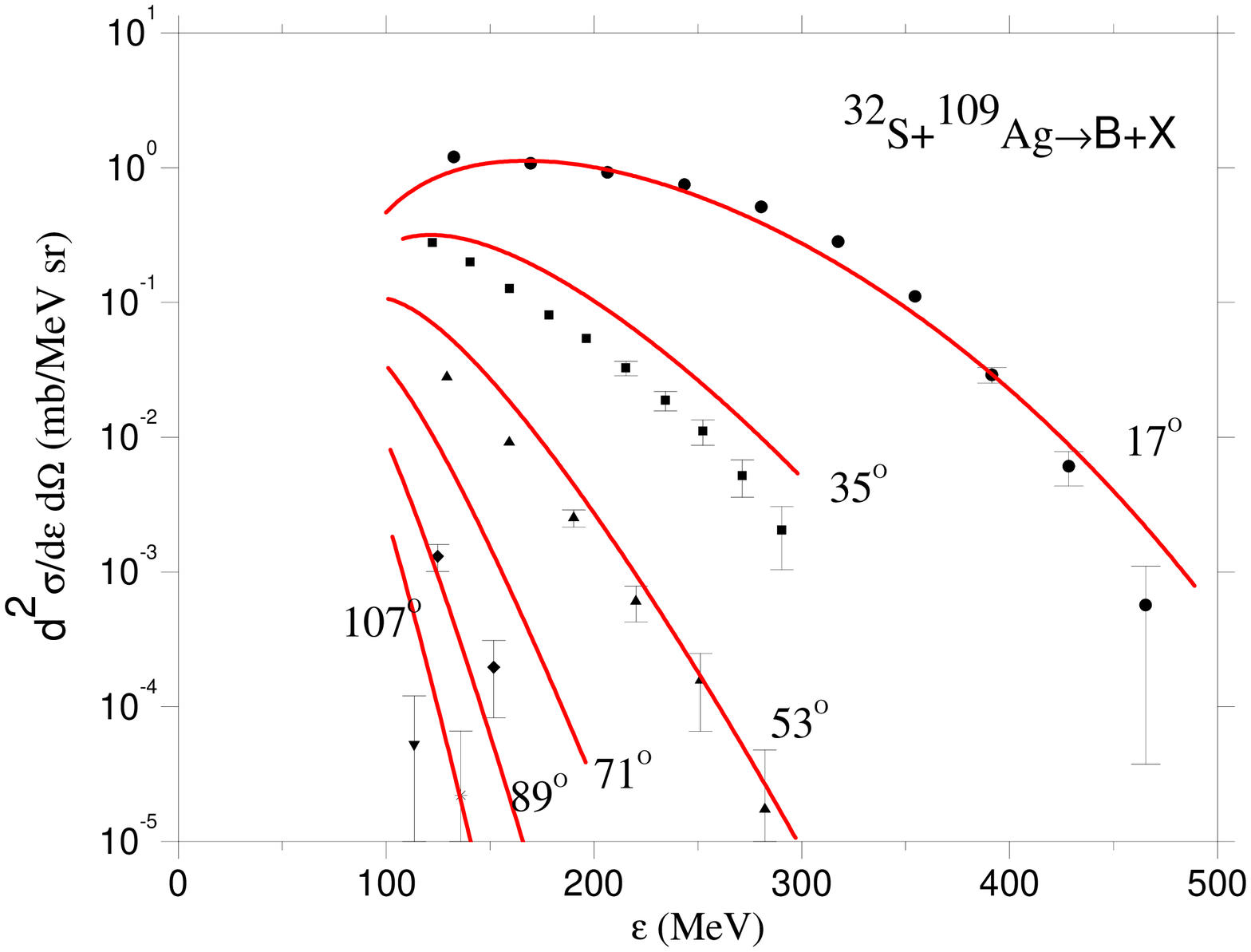}
\caption{(Colour online) Same as fig. \ref{Fig:V} but for the reaction $^{109}$Ag($^{32}$S,B)X.}
\label{Fig:Ag}
\end{center}
\end{figure}

\begin{figure}[h]
\begin{center}
\includegraphics[width=0.45\textwidth]{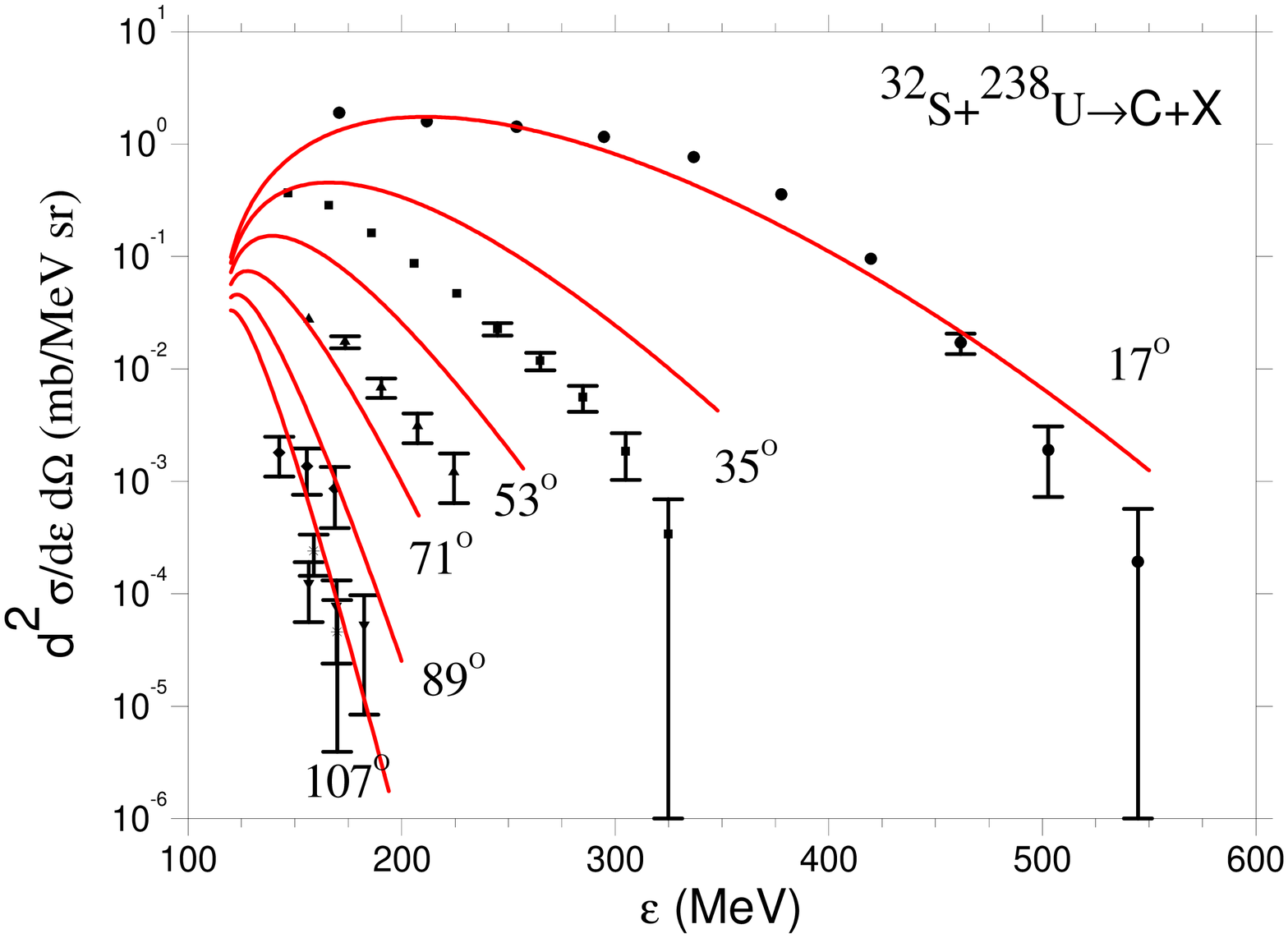}
\caption{(Colour online) Same as fig. \ref{Fig:V} but for the reaction $^{238}$Ag($^{32}$S,C)
X.}
\label{Fig:U}
\end{center}
\end{figure}

The quality of this procedure can be estimated from the comparisons with data. This is done in Figs. \ref{Fig:V}, \ref{Fig:Ag} and \ref{Fig:U}.
In producing the model spectra we have used this dependence to fit eq. (\ref{EQN_2}) to one spectrum at one angle. The model parameters were obtained by fitting the 17\degr spectrum. The temperature was assumed to follow a linear dependence $T(\varepsilon)=t_0+t_1\varepsilon$. The parameters $t_0$, $t_1$ and the normalisation $c$ where fitted.  Then the other spectra are predictions.

There is the general tendency of the calculations to overestimate the cross sections at the larger angles and high energies. However, it is just the energy range where $a(\varepsilon)$ could not be extracted because of lack of data.

We will now have a look to the derived functions $T$ and $E_0$.
\begin{figure}
\begin{center}
\includegraphics[width=0.45\textwidth]{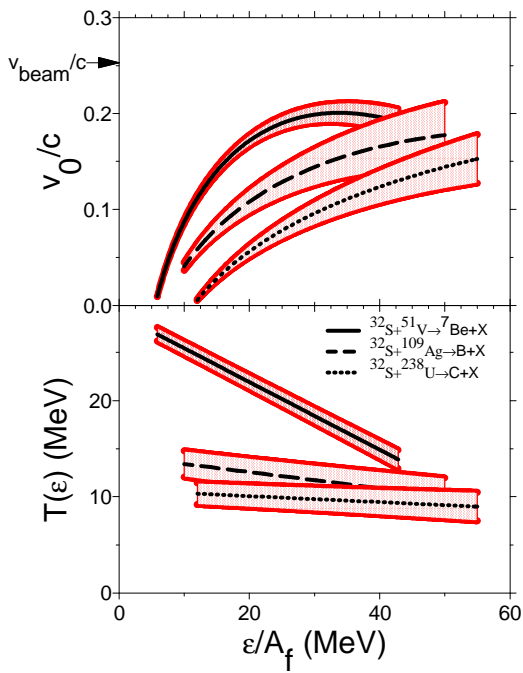}
\caption{(Colour online) The fitted function $T(\varepsilon )$ and the derived function $E_0(\varepsilon )$. The bands show the uncertainty of the functions.}
\label{Fig:T_E_0}
\end{center}
\end{figure}
These are shown in fig. \ref{Fig:T_E_0}. In addition to the energy dependence the error bands are shown.  The temperatures increase with decreasing fragment energy. That is what exactly is expected for an equilibrating system. Where as the energy dependence of the two heavy systems is modest it is much stronger for the lighter vanadium case. This can be understood in the Fermi gas model. Within this model is the thermal energy given by $U=A_{CN}/const*T^2$. For the same energy $U$ in a light system the temperature has to be large compared to a heavy system.

Also the source velocities behave as expected. For small energies the source velocities are small. For the highest energies they almost reach the beam velocity. This is similar to what was found in the moving source fits.

\subsection{Isotope ratios}\label{sub:Isotope-Ratios}
The models in the literature give energy spectra or total yields for different fragments. Here we will confront model predictions with isotopic cross sections with the goal to discriminate between different models. We, therefore, introduce shortly different models before comparing calculations with experiments.

\subsubsection{Quantum molecular dynamics}\label{sub:QMD}
The quantum molecular dynamics (QMD) model is closely related to the Boltzmann-Uehling-Uhlenbeck equation (also called Vlasov-Uehling-Uhlenbeck, Boltzmann-
Nordheim, or Landau-Vlasov equation) BUU/VUU approach~\cite{Aichelin91}. Scattering of nucleons is treated within a potential well. Collision and potential term are generated by the same bare interaction (which coincide in a classical theory), and the
Fermi statistics. The nucleons are approximated by Gaussian wave packets. The n-body density is a product of n Gaussians.

In the following we compare predictions of the model with isotopic cross sections.
\begin{figure}
\begin{center}
\includegraphics[width=0.45\textwidth]{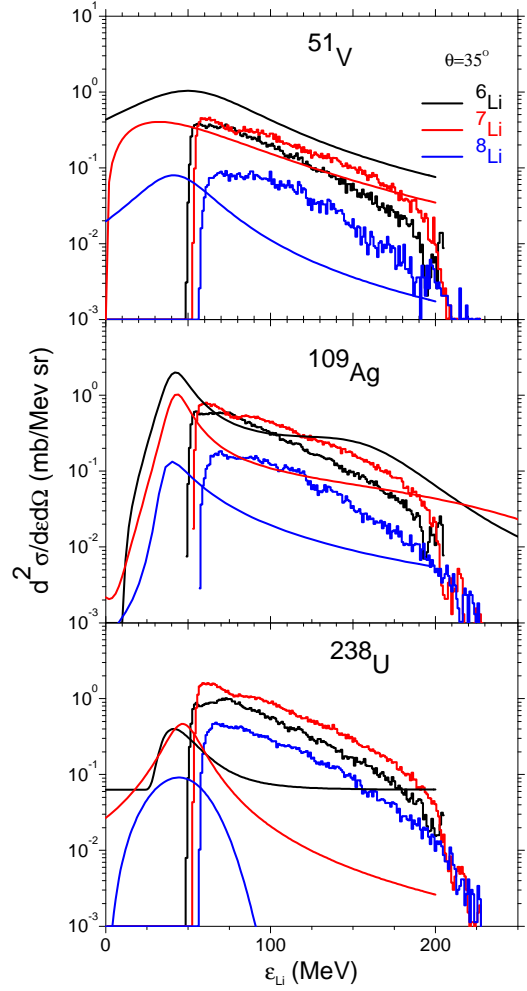}
\caption{(Colour online) Energy spectra for lithium emission from the interactions of sulfur with the indicate target nuclei. The experimental data are shown as narrow histograms while the PHITS calculations are shown as smooth curves. }
\label{Fig:Phits}
\end{center}
\end{figure}
The predictions of the QMD model were obtained with the code PHITS
\cite{Niita95, Phits06, Phits12}. In fig. \ref{Fig:Phits}
calculations for lithium isotope emission are plotted together with
the experimental energy spectra for 35\degr. The calculations
underestimate the experiment. The model predicts
$\sigma(A=6)>\sigma(A=7)>\sigma(A=8)$. However, the sequence in the
experiment is $\sigma(A=7)>\sigma(A=6)>\sigma(A=8)$. Unfortunately, PHITS does not allow to vary certain model parameters which might improve the theoretical results. A\"{o}so in Refs. \cite{Papa01} and \cite{Ono04} QMD calculations failed to reproduce experiments. This failure was overcome by obeying the fermionic nature of the nucleons.

\subsubsection{Coalescence model}\label{sub:Coalescence}
Another model is the coalescence model. Here we give a short description according to \cite{Gosset77}. We have neglected the index $f$ for the sake of simplicity.  We start from a nucleon momentum distribution $d^3M/dP^3$. The probability of finding a nucleon in a sphere in momentum space entered at momentum $P$ with radius $P_0$ is
\begin{equation}\label{eq:probability}
w=\left(\frac{4}{3}\pi P_0^3\right)\frac{1}{\bar{m}}\frac{d^3M}{dP^3}
\end{equation}
with $\bar{m}$ the average nucleon multiplicity. The probability of finding $n$ nucleons is then given by a binomial distribution. In case of small $\bar{m}$ as is the case in intermediate energy reactions it is reasonable to approximate the binomial distribution by a Poisson distribution. This yields
\begin{equation}
<w(n)>=(\bar{m}w)^ne^{\bar{m}w}/n! .
\end{equation}
Since $\bar{m}w$ is small the exponential can be ignored. The average probability to have $N$ neutrons and $Z$ protons on the coalescence sphere is then
\begin{equation}\label{eq:average_prob}
<w(Z,N)> = \frac{(\bar{m_z}w_Z)^Z}{Z!} \frac{(\bar{m_N}w_N)^N}{N!}.
\end{equation}
Insertion of eq. (\ref{eq:probability}) into this equation yields normalized to the coalescence volume \cite{Gosset77}
\begin{gather}
\frac{d^3M(Z,N)}{dP^3} = \frac{1}{Z!N!}(\frac{4}{3}\pi P_0^3)^{A-1}\nonumber \\\left[ \frac{d^3M(1,0)}{dP^3}\right]^Z \left[ \frac{d^3M(0,1)}{dP^3}\right]^N.
\label{eq:coalescence}
\end{gather}
The nucleon momentum density is related to the nucleon cross section by
\begin{equation}
\frac{d^3M}{dP^3} = \frac{1}{\sigma_0}\frac{d^3\sigma}{dP^3}
\label{eq:multiplicity}
\end{equation}
with $\sigma_0$ the total cross section. The relativistic invariant cross section is related to the non relativistic one by
\begin{equation}
\epsilon \frac{d^3\sigma}{dP^3}\simeq \frac{1}{P}\frac{d^2\sigma}{d\epsilon d\Omega}\,.
\end{equation}

Inserting this equation into eq. \eqref{eq:coalescence} gives
\begin{gather}\label{eq:Coalescence_final}
\frac{{d^2 \sigma _A (Z,N)}}{{d\varepsilon _A d\Omega }} = \frac{1}{{A\,\,Z!N!}}\left( {\frac{4}{3}\pi \frac{{P_0^3 }}{{\sigma _0 }}} \right)^{A - 1}\nonumber \\ \frac{1}{{(m\sqrt {2m\varepsilon } )^{A - 1} }}\,\,\left( {\frac{{d^2 \sigma (p)}}{{d\varepsilon d\Omega }}} \right)^Z \left( {\frac{{d^2 \sigma (n)}}{{d\varepsilon d\Omega }}} \right)^N
\end{gather}
with $m$ the nucleon mass, $m_A$ the fragment mass and $\varepsilon$ and $\varepsilon_A$ their energies. Here we have used $P_A=AP$. We further have ignored binding energy and Coulomb effects, i. e. $\epsilon_A=E\epsilon$ and $m_A=Am$ as well as relativistic effects. It was further assumed that the trajectories are not distorted by the Coulomb field.

\begin{figure}[h]
\begin{center}
\includegraphics[width=0.3\textwidth]{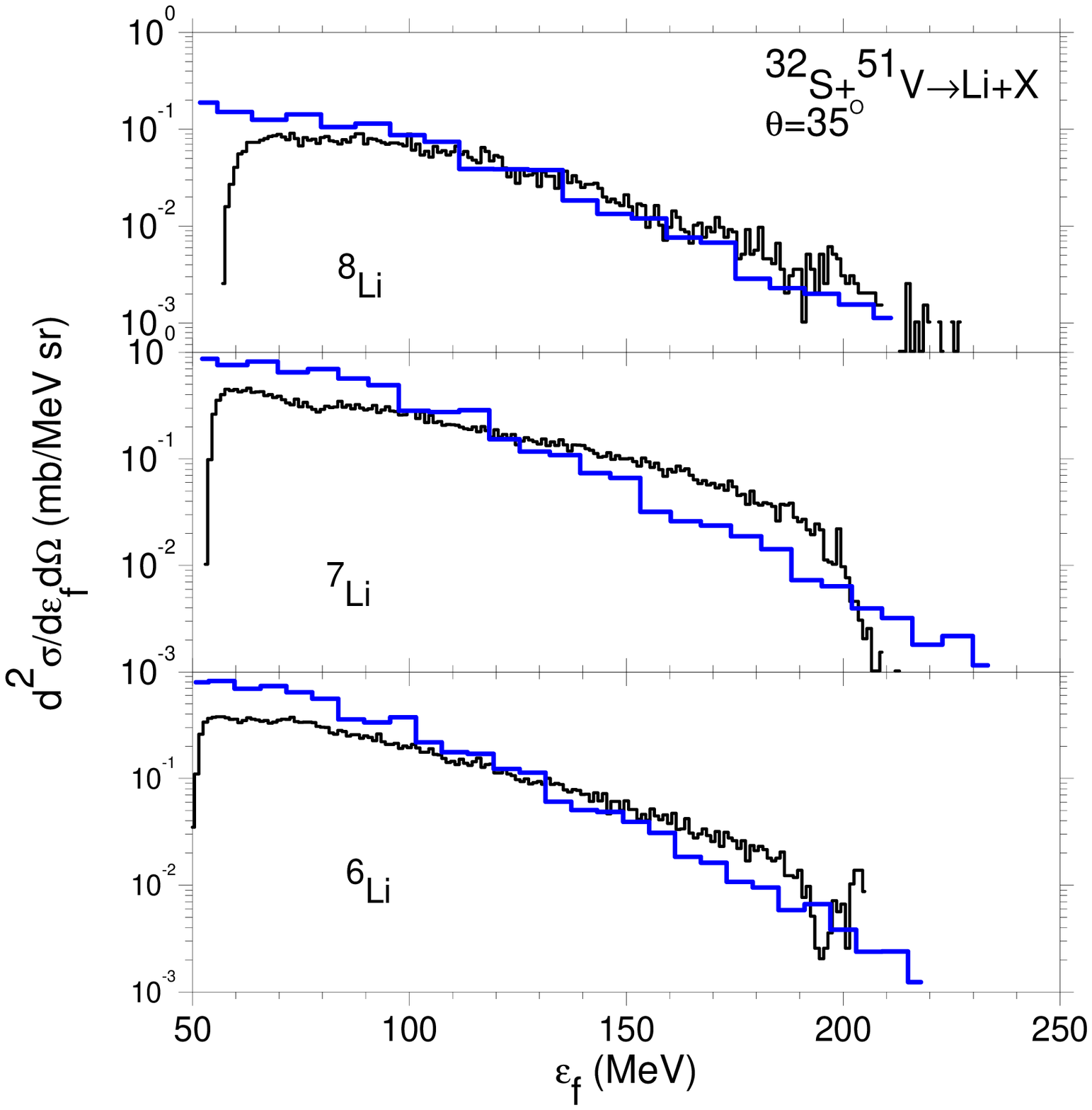}
\includegraphics[width=0.3\textwidth]{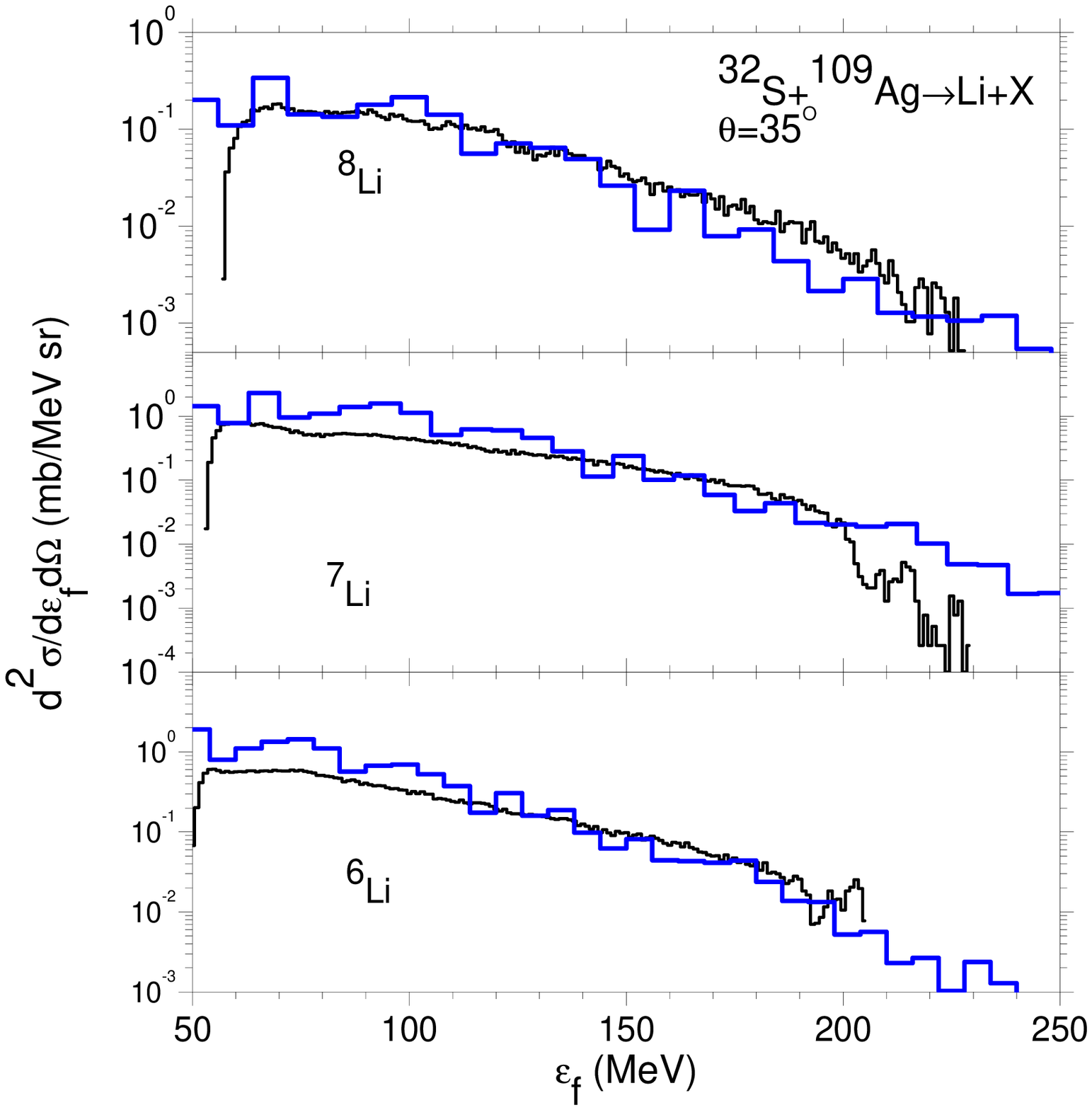}
\includegraphics[width=0.3\textwidth]{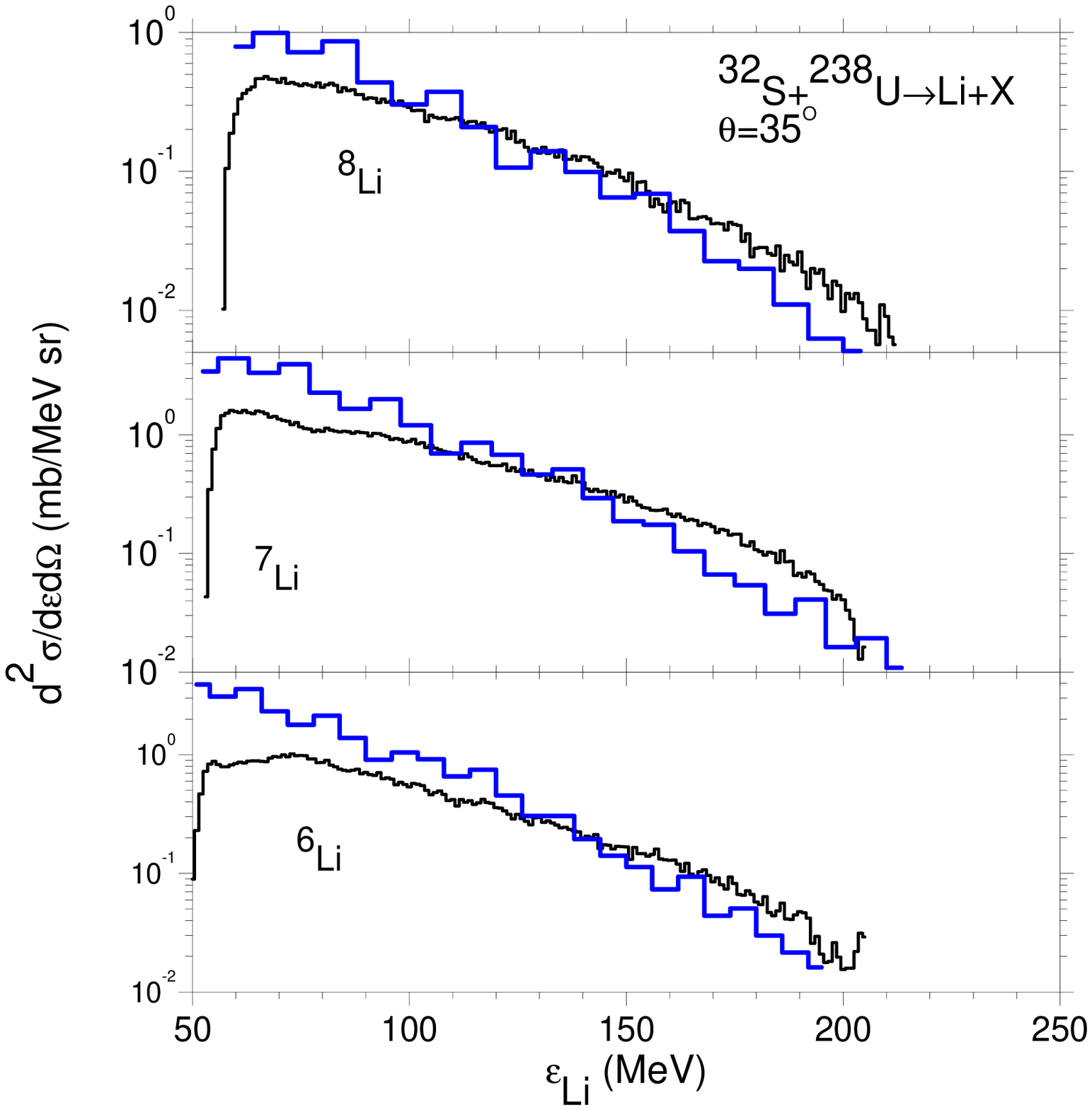}
\caption{(Colour online) Energy spectra of Lithium isotopes emitted from the intermediate systems of a target nucleus plus a sulfur nucleus. The experimental cross sections are shown as narrow histograms, the coalescence model calculations are shown as wide histograms. }
\label{Fig:Coal_LI}
\end{center}
\end{figure}

\begin{figure}[h]
\begin{center}
\includegraphics[width=0.4\textwidth]{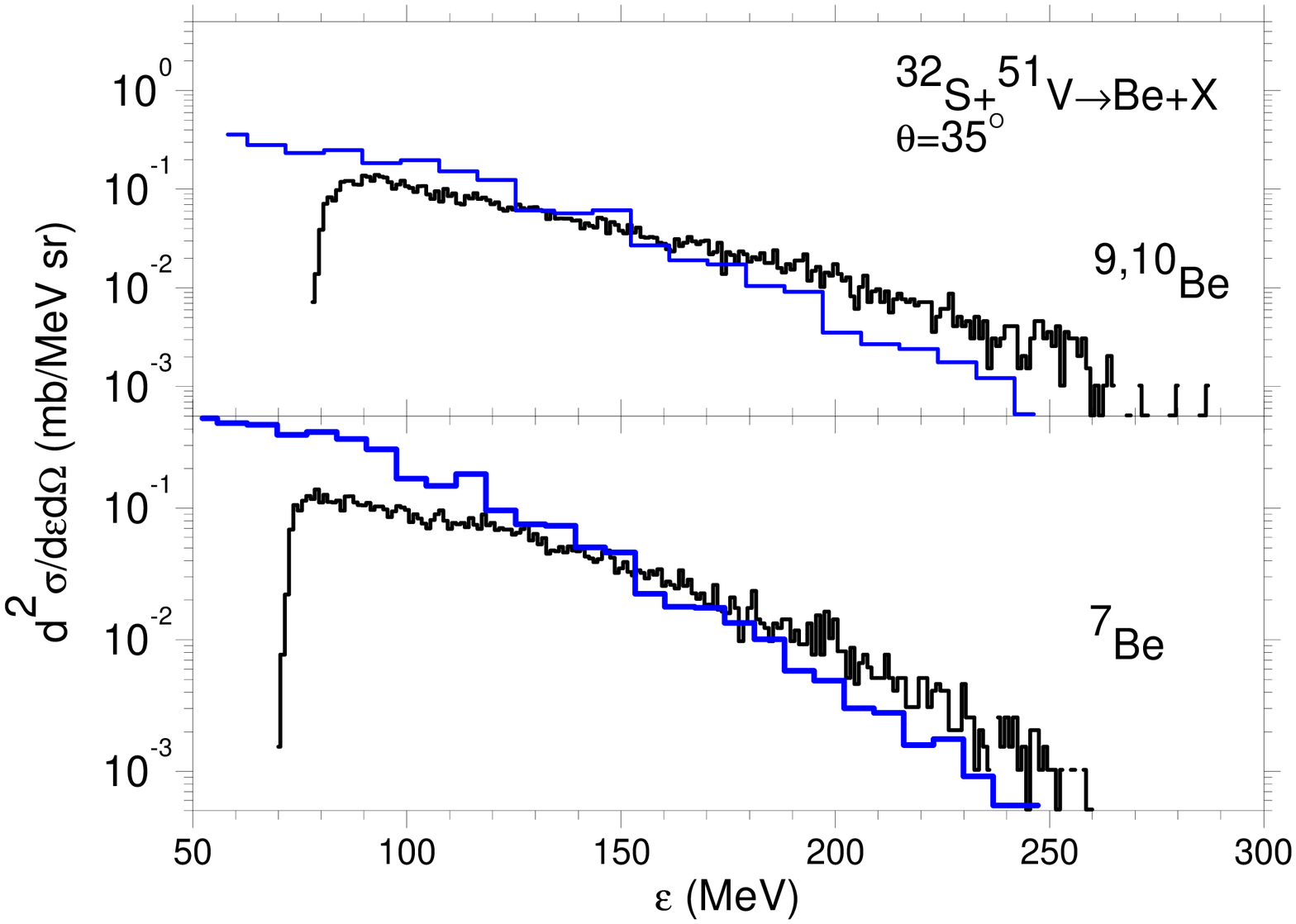}
\includegraphics[width=0.4\textwidth]{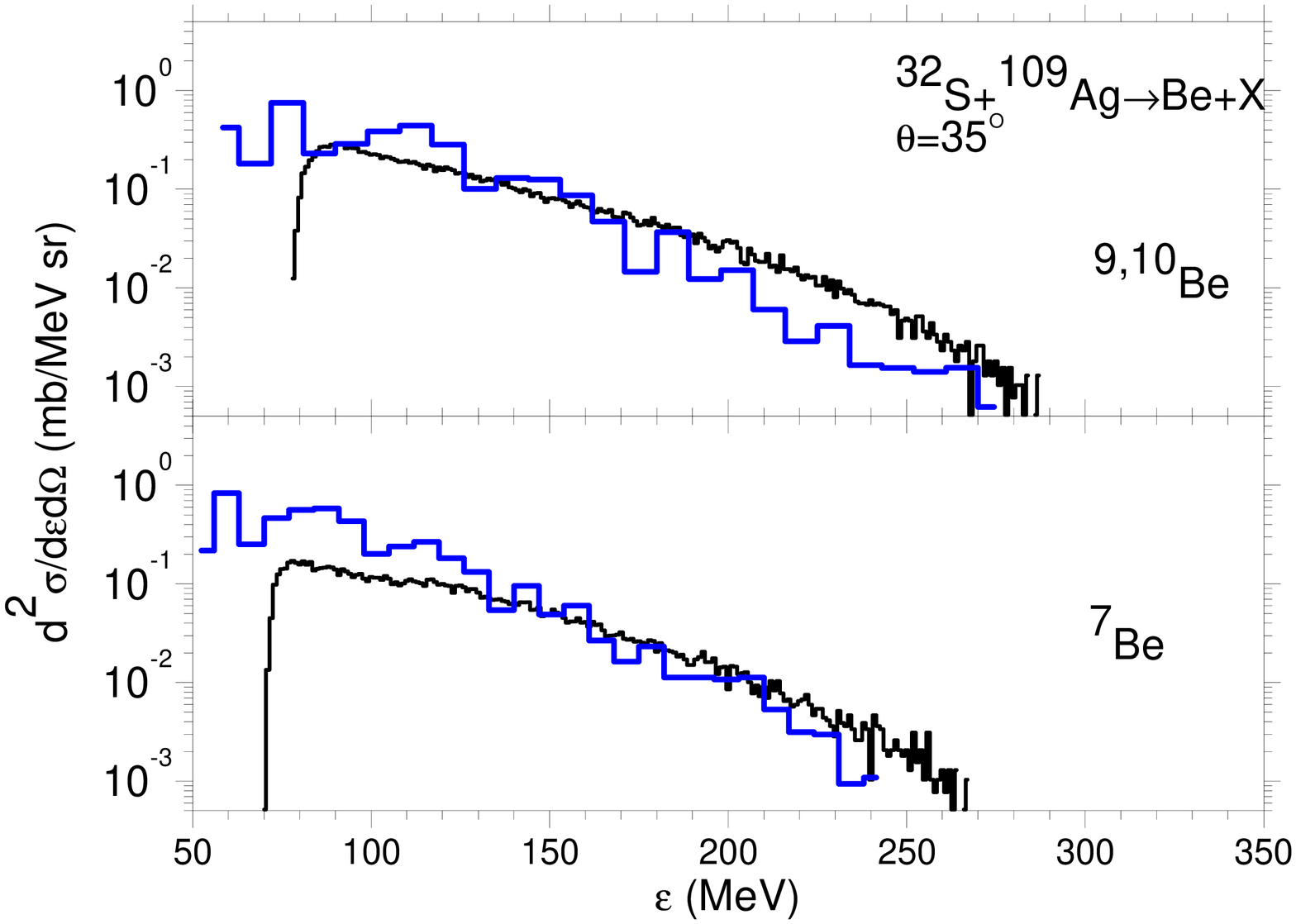}
\includegraphics[width=0.4\textwidth]{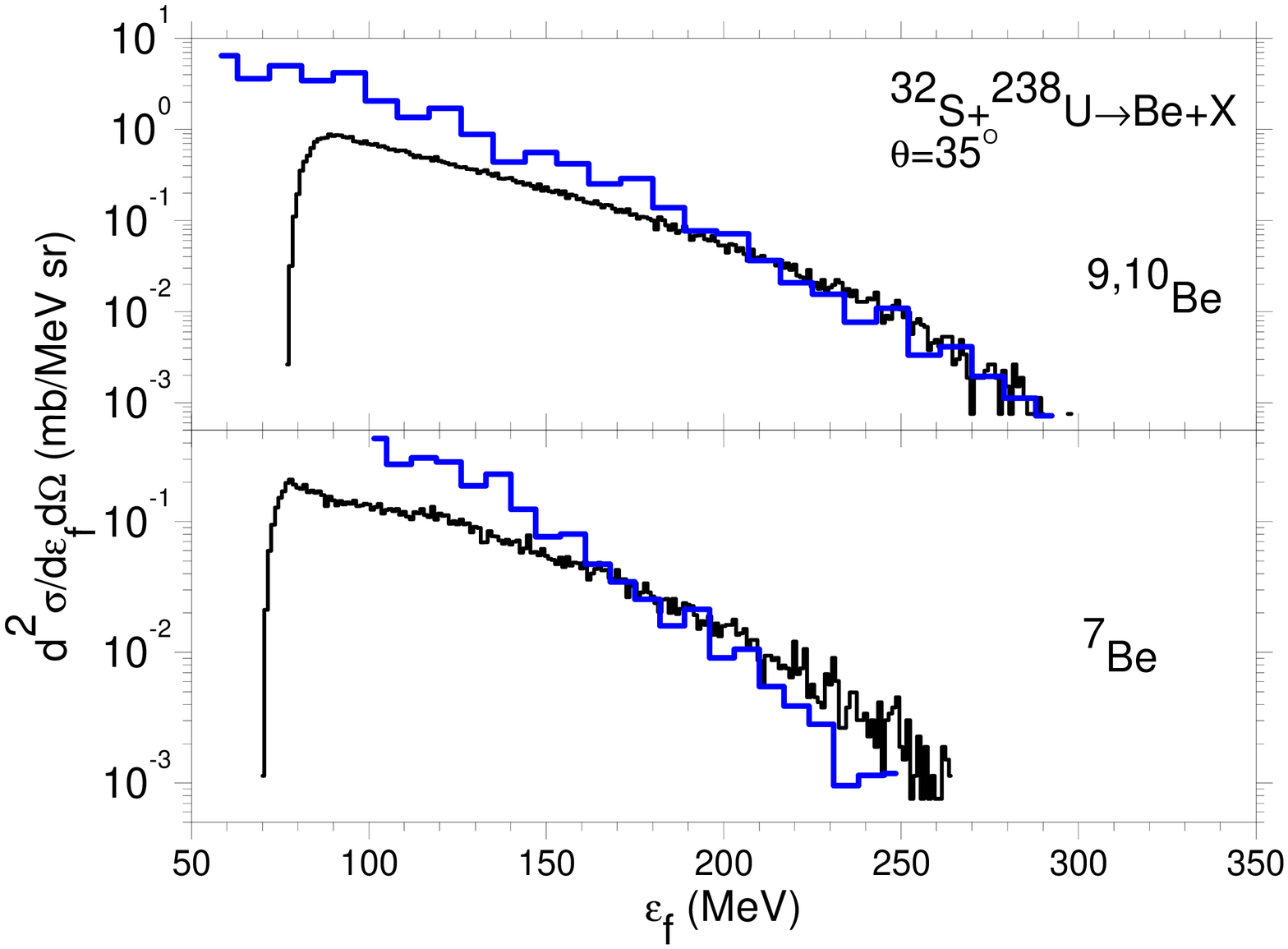}
\caption{(Colour online) Same as fig. \ref{Fig:Coal_LI} but for Beryllium isotopes.}
\label{Fig:Coal_Be}
\end{center}
\end{figure}

In almost all cases where this model has been applied to data analysis neutron spectra have not been available. In these cases the neutron yield per neutron in the composite target-projectile system was assumed to be proportional to the proton spectra per proton. Unfortunately in the present study even no proton spectra are available. We, therefore, produce both type of nucleon spectra from model calculations. The model used is again the QMD model in the programme PHITS. In \cite{Nakamura} neutron  energy spectra from heavy ion induced reactions were described by the PHITS calculations although on a double logarithmic scale.  For this purpose the evaporation option in the code is switched off. Another input for the coalescence model is the total reaction cross sections, which are calculated as $\sigma_0=\pi[1.2(A_T^{1/3}+A_P^{1/3})]^2$ (fm$^2$). For a comparison between data and calculations the forward angle of 35\degr was chosen. At more forward angles the cross section is dominated by projectile break-up while at backward angle evaporation from an equilibrates system dominates. Only the coalescence radii $P_0$ were adjusted by "eye" instead of $\chi$-square to reproduce the experimental cross sections. The model calculations are compared to the experiments in fig. \ref{Fig:Coal_LI} for the Lithium isotopes and in fig. \ref{Fig:Coal_Be} for the beryllium isotopes.

It seems that the calculated slopes are slightly steeper than the experimental ones. We will come back to this point.
\begin{figure}[h]
\begin{center}
\includegraphics[width=0.35\textwidth]{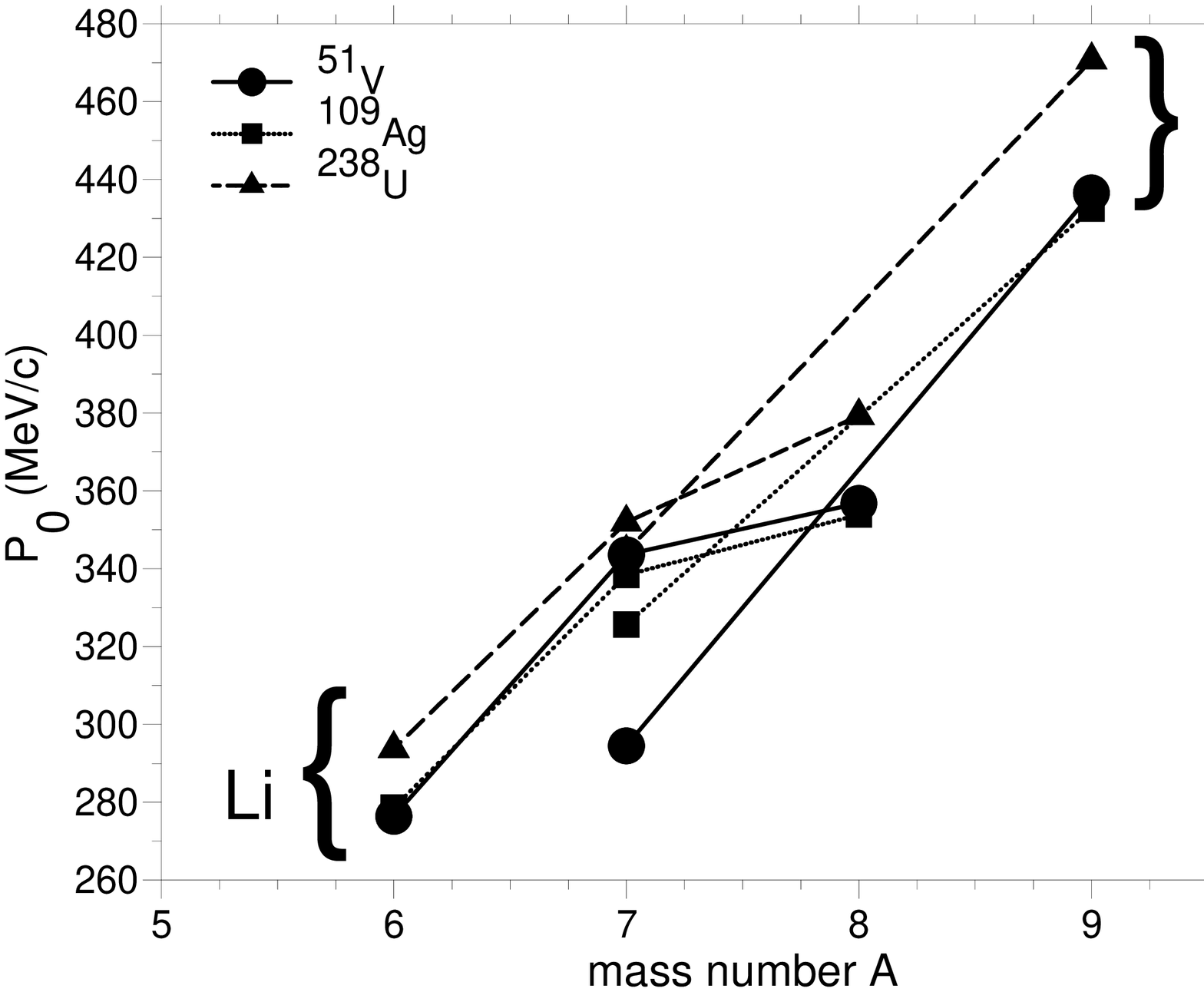}
\caption{The deduced coalescence radii for different elements as function of the fragment mass number $A$. The lines are to guide the eye.}
\label{Fig:P_0}
\end{center}
\end{figure}
The deduced coalescence radii are shown in fig. \ref{Fig:P_0} as function of the fragment mass. They increase almost linearly with the fragment mass. All values for {$^7$Li} are large than the arithmetic mean for {$^6$Li} and {$^8$Li}. This is an indication of the relative large abundance of {$^7$Li}. It is interesting to mention that the model has so far, to the best of our knowledge, been applied to isotopic resolved lighter fragments only up to $\alpha$-particles. In Ref. \cite{Cavinato94} cross sections for elemental resolved IMF yields \cite{Kim92} were compared with calculations. The deduced coalescence radii can therefore not be compared with the present ones. Furthermore a survival probability (assumed to be 0.4) was introduced in their formalism.

\begin{figure}
\begin{center}
\includegraphics[width=0.35\textwidth]{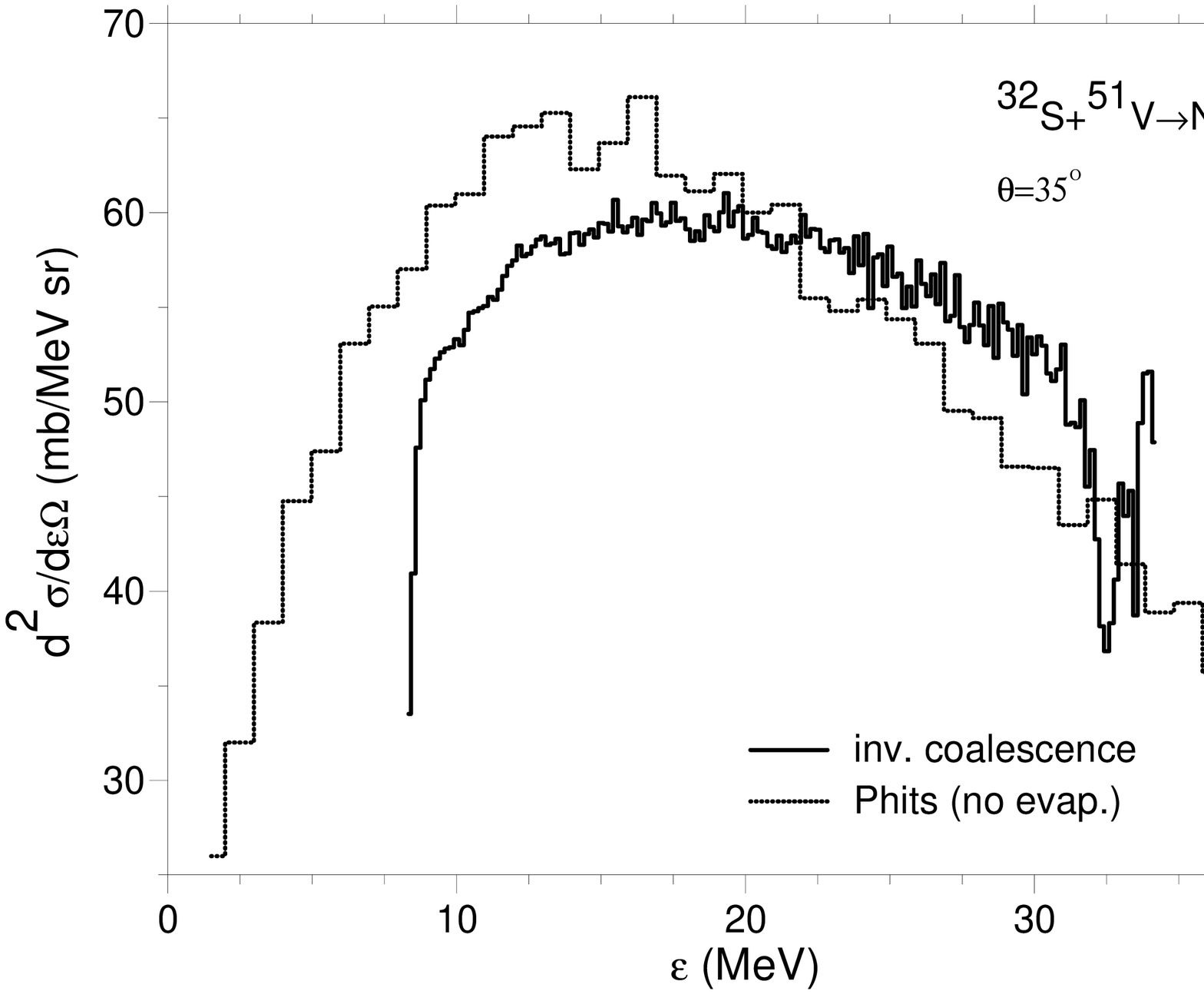}
\caption{Nucleon spectra. The one with small bin histogram is from the {$^6$Li} spectrum from the inverted coalescence model, the wide bin histogram from the QMD model calculation.}
\label{Fig:Nucleon_spectra}
\end{center}
\end{figure}

The too steep spectral slope of the model calculation visible in Figs. \ref{Fig:Coal_LI} and \ref{Fig:Coal_Be} is most probably due to a to steep slope of the proton and neutron spectra entering the calculation. In order to test this hypothesis we generate a nucleon spectrum from the {$^6$Li} spectrum by inverting the model eq. \eqref{eq:Coalescence_final} and replacing
$$\left( {\frac{{d^2 M(p)}}{{d\varepsilon d\Omega }}} \right)^Z \left( {\frac{{d^2 M(n)}}{{d\varepsilon d\Omega }}} \right)^N \to \left( {\frac{{d^2 M(N)}}{{d\varepsilon d\Omega }}} \right)^A.$$ The nucleon spectrum generated in this way is compared with the averaged proton-neutron spectrum from PHITS at 35\degr for the case of the vanadium target in fig. \ref{Fig:Nucleon_spectra}.
\begin{figure}
\begin{center}
\includegraphics[width=0.35\textwidth]{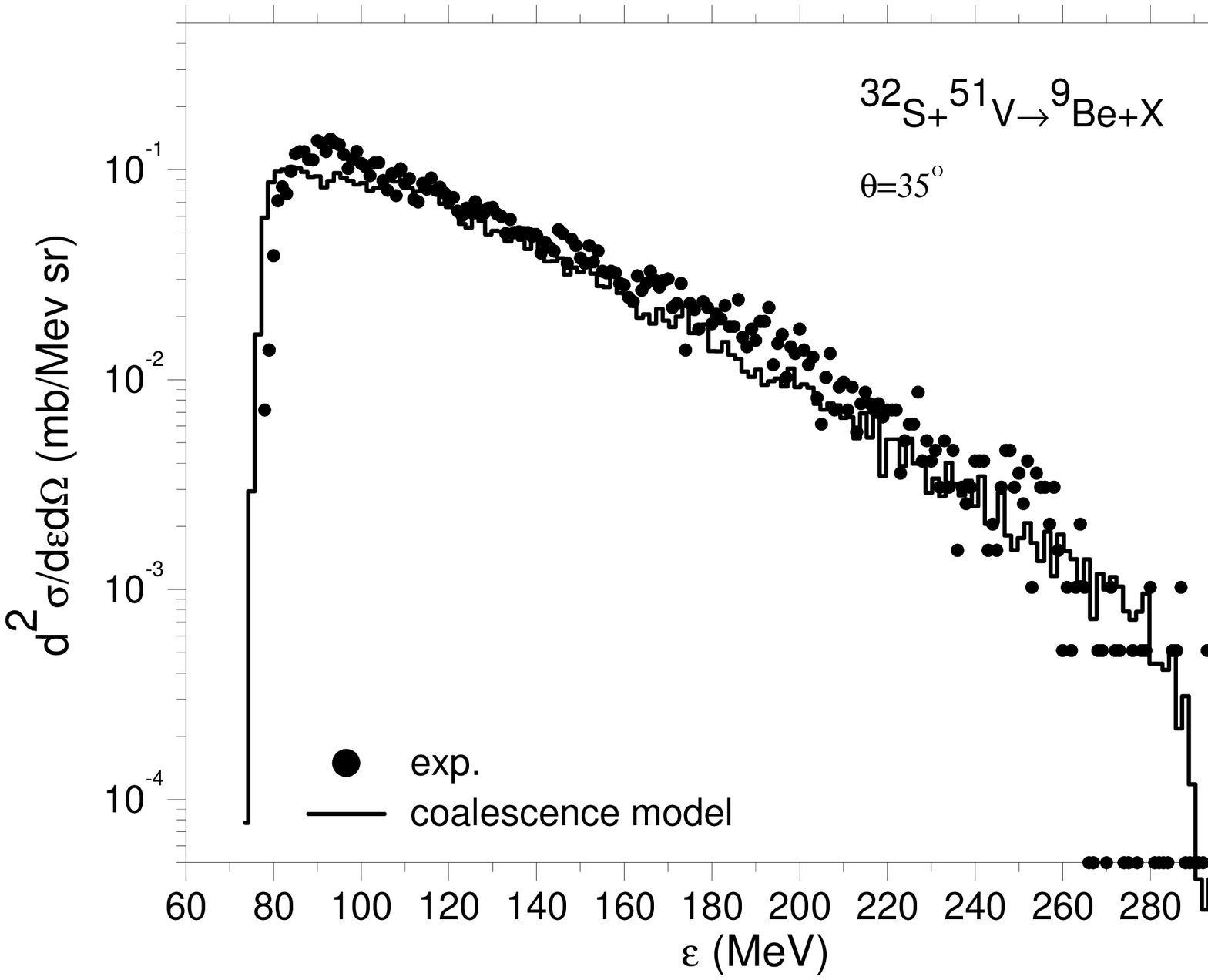}
\caption{Spectra for the reaction ${^{32}\text{S}}+{^{51}\text{V}}\to {^{9}\text{Be}}+X$. The experiment is shown by dots, the coalescence model with the nucleon spectrum from the inverted coalescence model is shown as histogram.}
\label{Fig:V_Be9_coal_inv}
\end{center}
\end{figure}
The shapes of the two spectra clearly differ. We now proceed and fold a fragment spectrum from the inverse coalescence spectrum.

For a comparison we chose the fragment being most different than the input fragment {$^6$Li} which is {$^9$Be}. This comparison is made in fig. \ref{Fig:V_Be9_coal_inv}. The coalescence radius is slightly changed to 113.6 MeV/c. However, one should keep in mind that the adjustments were made by eye as mentioned above. There is an almost perfect agreement between experiment and model calculation. This implies that the energy distribution of the ninth nucleon is not too different from the fifth neutron. The QMD spectra indicate a 10\% variation between both. However, its contribution is only 1/9 to the cross section, so the variation is on the 1\% level. The quality of the model calculation supports strongly the validity of the coalescence model.

As written in the introduction the density matrix formalism of Sato and Yazaki \cite{Sato81} takes the size of the emitted clusters explicitly into account. The coalescence volume is a folding between the spatial size of the emitting volume (hot spot or fireball, not defined) and the cluster wave function in coordinate space. Here we will follow a different approach to extract the size of the emitting source. The simplest approach would be to employ the phase space relation. Mekjian \cite{Mekjian77} has elaborated this simple phase space model assuming thermal and chemical equilibrium. Then the relation between coordinate space $V$ and momentum space $4/3 \pi P_0^3$ is
\begin{equation}
V=\left[ \frac{Z!N!A^3}{2^A}(2s_A+1)\exp(BE_{A}/T) \right]^{1/(A-1)}\frac{3h^3}{4\pi P_0^3}\,.
\label{eq:Mekjian}
\end{equation}
Here $2s_A+1$ is the spin degeneracy of the fragment, $EB_{A}$ its binding energy and $T$ its temperature. $h$ denotes Planck`s constant. We assume as in Ref. \cite{Lemaire79} $[e^{BE_{A}/T}]^{1/(A-1)} \approx 1$ and apply eq. \eqref{eq:Mekjian} to derive  coordinate space radii under the assumption $V=4/3\pi R^3$. Within this assumption the relation \eqref{eq:Mekjian} is a pure phase space relation and does not require thermal equilibrium.

\begin{figure}
\begin{center}
\includegraphics[width=0.45\textwidth]{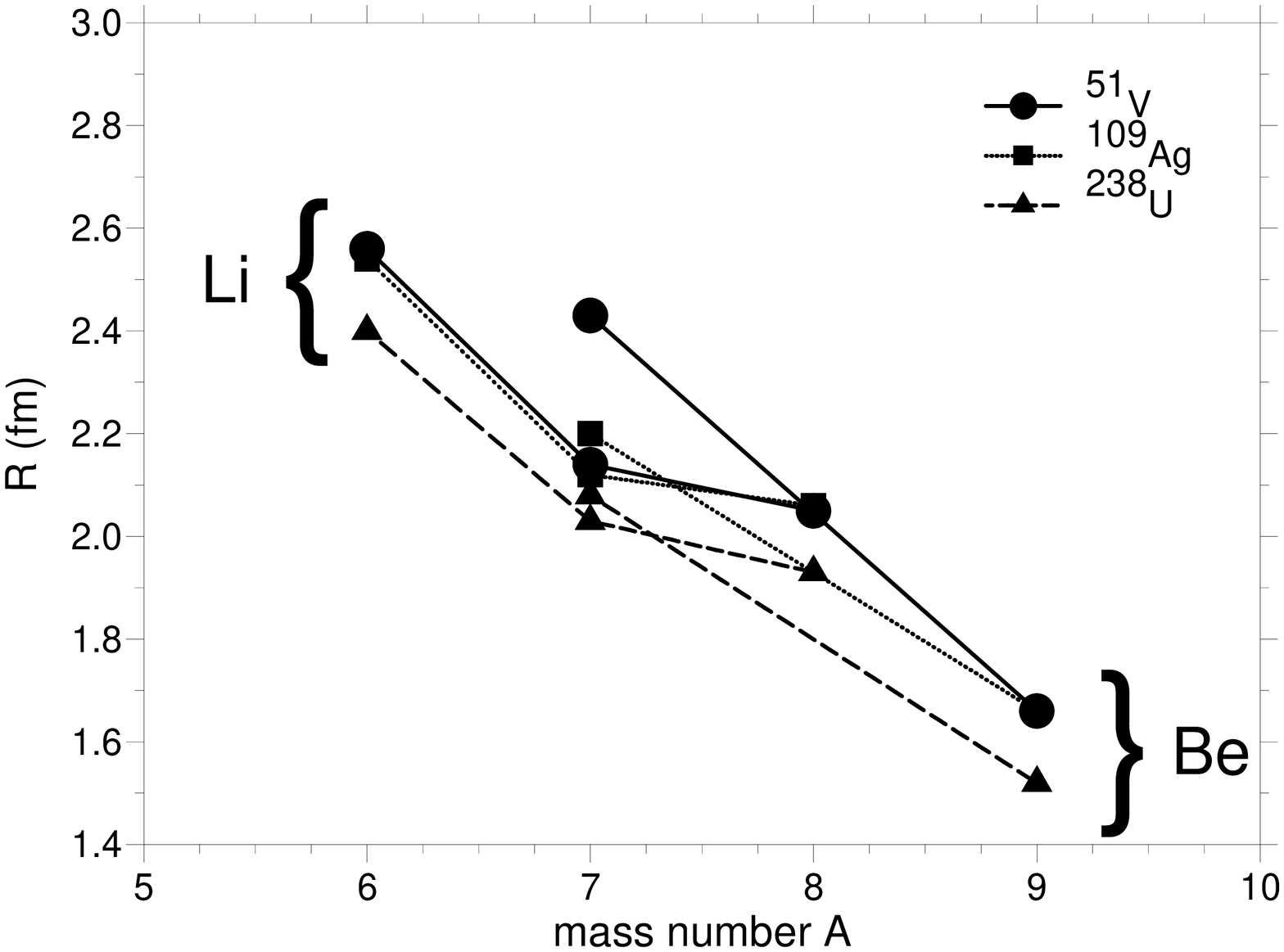}
\caption{Same as fig. \ref{Fig:P_0} but for radii in coordinate space.}
\label{Fig:Radii}
\end{center}
\end{figure}
The thus deduced radii $R$ are a linear dimension of the emitting volume and are shown in fig. \ref{Fig:Radii} as function of the fragment mass. This dependence is much stronger than the target mass dependence. This very weak dependence is an indication that emission occurs from an intermediate system which is smaller than a compound nucleus. It is interesting to note that the radii decrease with increasing fragment mass. However, this decrease is stronger than the one of the ground state radii \cite{Angeli13}.

\subsubsection{ Total elemental yields}\label{sub:Elemental_Yields}

As pointed out above, the GMSM is able to fit reasonably well the double differential cross sections. However, due to the finite thickness of the $\Delta$E counter the very low energy part of the energy spectra are missing and hence the deduced cross sections are lower limits (compare fig. \ref{Fig:Nucleon_spectra}).

\begin{figure}[h]
\begin{center}
\includegraphics[width=0.40\textwidth]{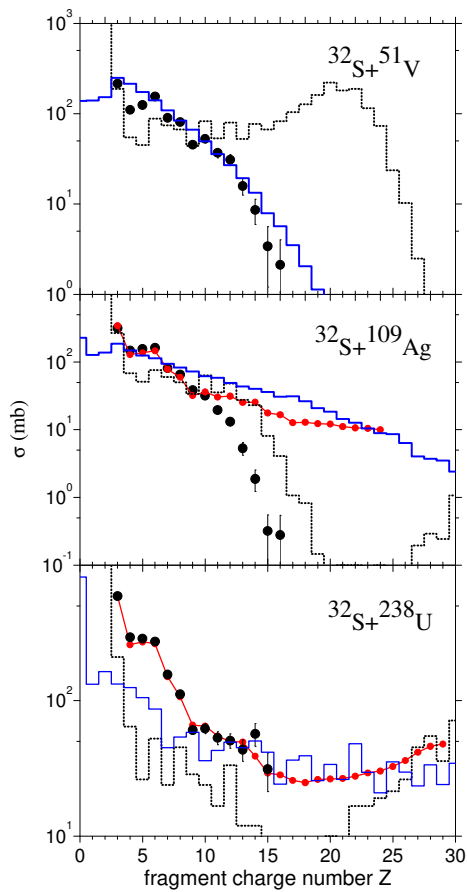}
\caption{(Colour online) The charge dependence of the elemental yields for the different target nuclei. The data for the present experiment are shown as full dots with error bars, those of Ref. \cite{Lleres87} as dots with connecting lines. The histograms (dotted lines) are the results from Phits calculations with QMD plus evaporation those with solid lines are SMM calculations. }
\label{Fig:Total_yields}
\end{center}
\end{figure}

The resulting distributions as function of fragment charge are shown in fig.~\ref{Fig:Total_yields}.
In all cases of the three targets the yield drops with increasing fragment charge. The cross section drops faster for $Z\gtrapprox 8$ for the vanadium target than for the uranium target. Such a behaviour might be due to the importance of fission in the latter case. Such an behaviour is also found in the data of \cite{Lleres87}. These data were obtained by integration over angular ranges from 30\degr - 110\degr in the case of silver and 30\degr - 70\degr in the case of a gold target, which we compare here with uranium. There is a remarkable plateau formed for Be, B, and C. This is also visible in the data of Lleres et al. \cite{Lleres87}. One reason is the sharp drop due to the nonexistence of a bound $^8$Be isotope. For Carbon the cross section might be large due to the favourable $Q$-value for binary break up of the beam particles:
\begin{center}
\begin{tabular}{cc}
${^{32}\text{S}}\to {^{10}\text{B}}+{^{22}\text{Na}}$: & -33.2\text{MeV}\\
${^{32}\text{S}}\to {^{12}\text{C}}+{^{20}\text{Ne}}$: & -19.3\text{MeV}\\
${^{32}\text{S}}\to {^{14}\text{N}}+{^{18}\text{F}}$: & -30.0 \text{MeV}\\ 
\end{tabular}
\end{center}

While the agreement between the two data sets is good for the heavy target it fails in the case of silver for fragment yields with $Z>9$. The reason for this is unclear.

We have used the programme PHITS \cite{Phits06}, \cite{Phits12} to calculate the charge distribution for the three reactions employing the QMD plus evaporation approach. The results are shown as histograms in fig. \ref{Fig:Total_yields}. The trends in the case of the silver and uranium target are reproduces although not on a quantitative level. In the case of the lightest target vanadium the calculation shows an enhancement of the  cross sections around $\approx 20$ which is not in the data.

It is worth mentioning that the coalescence model, as it is used here, does not account for excited fragments which then decay not only by $\gamma$ emission but also be particle emission. This may influence the spectra of neighbouring lighter fragments. A detailed study in the case of excited $^{10}$C fragments is reported in \cite{Greinier08}.

\subsubsection{Statistical multifragmentation model}
 The statistical multifragmentation model (SMM) \cite{Bondorf} \cite{Gross97} assumes that the thermalised residual nucleus of the first stage of the projectile nucleus collision undergoes a statistical breakup. At first the nucleus expands to a certain volume and then breaks up into nucleons and hot fragments. All possible breakup channels are considered. The probability $w_j$ of a specific decay channel $j$ of the nucleus excited to the energy $E^*$ is proportional to the exponential function of the entropy $S_j (E^*)$, which (besides the excitation energy) depends also on other parameters of the system: $w_j\propto\exp[S_j (E^*)]$. The model treats the formation of a compound nucleus as one of the decay channels. This allows for the transition from evaporation at low energies to multifragmentation at high excitations on the basis of the available phase space. It is assumed that at the breakup time the nucleus is in thermal equilibrium characterised by the channel temperature T . The light fragments with mass number $A<4$ and atomic number $Z<2$ are treated as structureless particles, i.e., they have only translational degrees of freedom. The heavier fragments are considered as heated drops of nuclear liquid, thus their individual free energies are parameterised according to the liquid-drop model, i.e., they are equal to sum of the bulk, surface, Coulomb, and symmetry energies. The temperature dependencies of these parameters are compiled in Ref. \cite{Donangelo01} together with a code description. Since the symmetry energy is small its temperature dependence is ignored.

 Contrary to Gross et al. \cite{Gross97} who employed microcanonical ensembles in the present formulation macrocanonical ensembles were used. Whereas the term channel corresponds to a member of a final state this is named partition. The constraints are mass and charge conservation:
 \begin{gather}\label{eq:conservation_A}
 A_f=\Sigma _{A,Z} N(A,Z)A \\ \label{eq:conservation_Z}
 Z_f=\Sigma_{A,Z} N(A,Z) Z\,,
 \end{gather}
with $N(A,Z)$ the multiplicity of the isotope with mass number $A$ and charge $Z$. The distribution of partition probabilities in the macrocanonical approximation is given by:
\begin{equation}\label{eq:distr_partition}
W_f=exp(-\Omega_f/T)/\Sigma_{f'} exp(-\Omega_{f'}/T)
\end{equation}
with
\begin{equation}\label{eq:thermodynamical}
\Omega_f = F_f-\mu A_f-\mu_Z Z_f\,.
\end{equation}
the thermodynamical potential. The $\mu$'s are the chemical potentials which were found from  \eqref{eq:conservation_A} and \eqref{eq:conservation_Z}. The averaging is necessary since the energy fluctuates form partition to partition. $F$ is the free energy.

We have used this code to calculate fragment charge distributions for the three systems. The total excitation energy $E^*$ was calculated assuming full momentum transfer from the projectile to the emitting source. The results of these calculations are also shown in fig. \ref{Fig:Total_yields}. We have multiplied the results by 250 for the two lighter target nuclei and by 200 for the uranium case. For the vanadium case the calculation agrees much better with the experimental results than the QMD calculations. In the case of the silver target the present results start to differ from those from \cite{Lleres87} from neon on. For the heavier fragments the missing low energy yield becomes more important. The SMM calculation follows the Lleres result, again contradictory to the QMD calculation. For uranium the Lleres result although for gold and the present results agree with each other because the maximum in the spectra for this case is at higher energies than for the silver target. The agreement between both data sets and the SMM calculation is good for the heavy fragments and similar to the QMD calculation for the lighter fragments. In general SMM results are favourable to QMD results.

\par
\medskip
\noindent \section{ Summary and conclusion}\label{sec:Summary-Conclusions}

We have studied inclusive emission from IMF's following the
bombardment of $^{51}V$, $^{109}Ag$, $^{197}Au$,
and $^{238}U$ target nuclei with $^{32}S$ ions at an energy of 31.6 MeV A. The
measurements were performed with two
large solid angle telescopes in a range from 17 degrees to 107 degrees.

The energy spectra of lighter fragments are smooth curves with angular distributions peaked into the beam direction. At the most forward measuring angle a weak projectile-like component was found. In the case of uranium there is a second component for heavy fragments which is not visible for the other nuclei studied. This component may be attributed to quasi-free scattering with nucleon exchange while the first component may be due to decay from an excited projectile-like system. Also the long ranged Coulomb force may be responsible for this component.

The energy spectra were analysed in terms of the GMSM. The energy dependence of the model parameter $a(\varepsilon )$ is similar to the one in previous studies~\cite{Buh92} for not to high emission energies..

Deduced charge distributions - differential as well as elemental ones - show more an exponential slope than a power law dependence.

The energy dependencies of relative isotopic yields are smooth for lithium isotopes but show a strong energy and mass dependence in the case of beryllium isotopes. The yield for $^7$Li is larger than the one for the other isotopes. The reason is at the moment not clear. The same behaviour is found in the data from refs. \cite{Mac85a} for a heavy system and less pronounced in the data \cite{Buh92} for lighter systems. The yields for $^6$Li and $^7$Li are the same for proton induced reactions with a similar total beam energy for a light system \cite{Fidelus14} (aluminum) and heavier systems (nickel and gold) \cite{Budzanowski10}. For even much higher proton beam energies the behaviour of the intensities is as in the present case \cite{Hirsch84}.

Isotopic yields were very well described by the coalescence model. This model needs only one adjustable parameter. Such quality of reproduction could not be obtained by QMD model calculations. Contrary to these dynamical models is SMM which is statistical. However, although the physical picture is quite different than the coalescence model, it also yields a satisfactory description of charge distributions.

\centering
\textbf{Acknowledgement}
We are grateful to the accelerator crew for providing the excellent beam. Discussions with F. Goldenbaum and A. Magiera were found very helpful. Two of us (H.M. and T.K.) thank the IKP of FZ J\"{u}lich for hospitality.




\end{document}